\documentclass[journal]{IEEEtran}
 \usepackage{cite}

\ifCLASSINFOpdf
   \usepackage[pdftex]{graphicx}
\usepackage{epstopdf}
\else
   \usepackage[dvips]{graphicx}
\fi
\usepackage{latexsym}
\usepackage{amsmath}
\usepackage{multirow}
\usepackage{bm}
\usepackage{algorithmic}
\usepackage[titlenumbered,ruled]{algorithm2e}
\usepackage{balance}
\usepackage[hyphens]{url}
\usepackage{dsfont}
\usepackage{subfig}
\usepackage{color}
\usepackage{multirow}
\usepackage{longtable}
\usepackage{supertabular}

\usepackage[english]{babel} 
\usepackage{blindtext}
\usepackage{bookmark}
\bookmarksetup{numbered}

\newcommand{\subparagraph}{}

\usepackage{tikz}
\def\checkmark{\tikz\fill[scale=0.4](0,.35) -- (.25,0) -- (1,.7) -- (.25,.15) -- cycle;} 
\usepackage{amssymb}
\usepackage{pifont}

\usepackage{array,booktabs}
\usepackage{cleveref}

\newenvironment{Blue-Color}{\par\color{blue}}{\par}
\newenvironment{Red-Color}{\par\color{red}}{\par}

\begin{document}

\title{\vspace{-.5cm}A Survey on Multi-AP Coordination Approaches over Emerging WLANs: Future Directions \\ and Open Challenges}

\author{Shikhar Verma,~\IEEEmembership{Member,~IEEE,}
Tiago Koketsu Rodrigues, ~\IEEEmembership{Member,~IEEE,}
	Yuichi Kawamoto,~\IEEEmembership{Member,~IEEE,}
	Mostafa M. Fouda,~\IEEEmembership{Senior Member,~IEEE,}
	and~Nei Kato,~\IEEEmembership{Fellow,~IEEE}
	\thanks{S. Verma, , T.K. Rodrigues, Y. Kawamoto, and N. Kato are with the Graduate School of Information Sciences, Tohoku University, Sendai, Japan. Emails: 
		\{shikhar.verma, tiago.gama.rodrigues, youpsan, and kato\}@it.is.tohoku.ac.jp}
  \thanks{M.M. Fouda is with the Department of Electrical and Computer Engineering at Idaho State University, ID, USA.  Email: mfouda@ieee.org}
}

\maketitle
\begin{abstract}
Recent advancements in wireless local area network (WLAN) technology include IEEE 802.11be and 802.11ay, often known as Wi-Fi 7 and WiGig, respectively. The goal of these developments is to provide Extremely High Throughput (EHT) and low latency to meet the demands of future applications like as 8K videos, augmented and virtual reality, the Internet of Things, telesurgery, and other developing technologies. IEEE 802.11be includes new features such as 320 MHz bandwidth, multi-link operation, Multi-user Multi-Input Multi-Output, orthogonal frequency-division multiple access, and Multiple-Access Point (multi-AP) coordination (MAP-Co) to achieve EHT. With the increase in the number of overlapping APs and inter-AP interference, researchers have focused on studying MAP-Co approaches for coordinated transmission in IEEE 802.11be, making MAP-Co a key feature of future WLANs. Moreover, similar issues may arise in EHF bands WLAN, particularly for standards beyond IEEE 802.11ay. This has prompted researchers to investigate the implementation of MAP-Co over future 802.11ay WLANs. Thus, in this article, we provide a comprehensive review of the state-of-the-art MAP-Co features and their shortcomings concerning emerging WLAN. Finally, we discuss several novel future directions and open challenges for MAP-Co.

\end{abstract}

\begin{IEEEkeywords}
Multi-AP Coordination, Wireless local area network, IEEE 802.11ay, IEEE 802.11be, millimeter Wave. 
\end{IEEEkeywords}

\IEEEpeerreviewmaketitle

\section{Introduction}
 \begin{table}[!ht]
	\renewcommand{\arraystretch}{1.3}
	\caption{List of Acronyms.}
	\label{acronym}
	\centering

	\begin{tabular} {| p{2.5cm} | p{5.3cm} | }
	
		\hline ~~~~~~~~~\textit{Acronym} 				&~~~~~~~~~~~~~~~~\textit{Definition} 
		\\ 
		\hline

C-OFDMA	&Coordinated Orthogonal frequency-division multiple access \\ \hline
CB	&Compressed beamforming \\ \hline
CBF	&Coordinated beamforming \\ \hline
CCA/CS	&Clear channel assignment/carrier sense\\ \hline
CSI	&Channel state information \\ \hline
CSMA/CA	&Carrier sense multiple access/collision avoidance \\ \hline
CSR	&Coordinated spatial reuse  \\ \hline
D-MIMO	&Distributed MIMO \\ \hline
EHT 	&Extreme high throughput  \\ \hline
ICI	&Inter-cell interference \\ \hline 
IRS	&Intelligent reflecting surface  \\ \hline
JTX	&Joint Transmission \\ \hline
MA	&Master AP \\ \hline
MAC	&Medium Access Control \\ \hline
MAP-Co 	&Multi-Access Point Coordination \\ \hline
MC 	&Master controller \\ \hline
MCSR	&MAC-layer based CSR \\ \hline
MU-MIMO	&Multi-user multi input multi output \\ \hline
NAP	&Neighboring slave APs \\ \hline
NDP	&Null data packet \\ \hline
NDPA	&Null data packet announcement \\ \hline
OBSS	&Overlapping basic service sets \\ \hline
OBSS-PD	&Overlapping basic service sets-packet detect \\ \hline
OPCSR	 &OBSS PD-based CSR \\ \hline
PCSR	&Parameterized CSR \\ \hline
RSSI	&Receiving signal strength index \\ \hline
RU	&Resource units  \\ \hline
SIFS	&Short interframe space \\ \hline
SINR	&Signal-to-Interference-plus-Noise Ratio  \\ \hline
SNR	&Signal-to-noise-ratio   \\ \hline
TP	&Transmission power  \\ \hline
TPC	&Transmission power control   \\ \hline
TXOP	 &Transmission opportunities  \\ \hline
	\end{tabular}
	
\end{table}
\begin{figure*}
	\centering
	\includegraphics[width=7in]{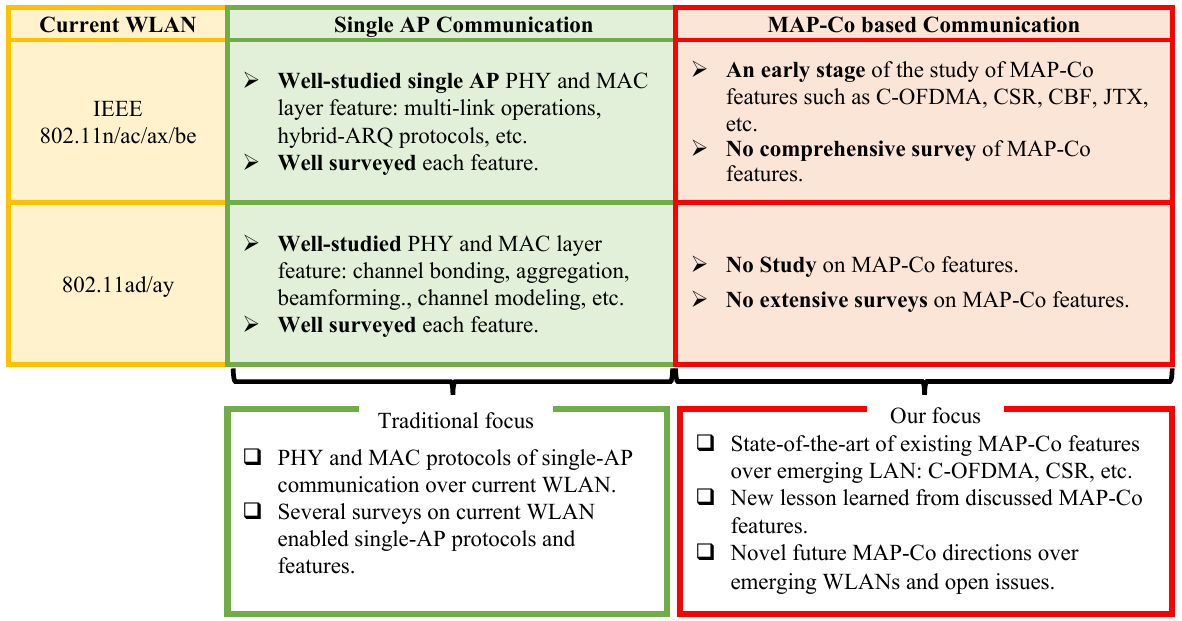}
	\caption{Comparison of studies in current WLAN enabled single-AP and multi-AP coordinated communication, presenting comparison between traditional and our focus in this paper.}
	\label{feature-comparison}
	\vspace{-0.5cm}
\end{figure*}

\IEEEPARstart{O}{ver} the last two decades, wireless local area networks (WLAN) have treaded the path of continuous evolution to provide high data rates at lower costs~\cite{shikhar}. In this regard, Wi-Fi Alliance with IEEE has proposed IEEE 802.11ax (Wi-Fi 6), and IEEE 802.11ad to provide extremely high throughput (EHT)~\cite{ieee1}. Especially, post-pandemic, business models are dramatically switching towards digitalization of everything, drastically increasing the number of connected devices and traffic over WLAN. Moreover, users will soon experience a quantum leap in new applications such as 8K videos, virtual/augmented reality (VR/AR), and large-scale connected sensors, leading to new advanced network requirements~\cite{vr}. For instance, VR/AR will require EHT (20 gigabits per second (Gbps)) with low latency (lower than 5ms) and stringent reliabilities. Meeting such extremely high requirements is beyond the capacity of 802.11ax and 802.11ad. In this regard, IEEE has amended 802.11ax and 802.11ad to propose new standards 802.11be (namely Wi-Fi 7) and 802.11ay (namely WiGig), respectively~\cite{kho,ieee2}. In this article, we use 802.11be and Wi-Fi 7 interchangeably as we use 802.11ay and WiGig interchangeably. 
In 2018, IEEE 802.11 approved the creation of a new task group for next-generation IEEE 802.11be (TGbe) to define physical (PHY) and medium access control (MAC) protocols for enabling at least 30 Gbps throughput while ensuring backward compatibility~\cite{lopez}. The task group also focuses on reducing worst-case latency and jitters to support time-sensitive applications such as AR/VR. In PHY layer amendments, 802.11be added new bandwidths such as continuous 240 MHz, continuous 320 MHz, and noncontinuous 160+160 MHz, multi-user (MU) resource unit (RU) assignment, 4096-quadrature amplitude modulation, permeable formats, and puncturing techniques~\cite{kho}. In the MAC layer, multi-link operation, 16 spatial streams MU-MIMO, and Hybrid automatic repeat request are the main discussed features. Similarly, from 2015 to 2021, IEEE 802.11 worked on amendments of 802.11ad to provide at least 20 Gbps throughput and latency lower than 5 ms over 60GHz while maintaining power efficiency~\cite{ieee2}. 802.11ay proposed channel bonding and aggregation of 2.16 GHz channels, multi-user multi input multi output (MU-MIMO) beamforming, and MIMO multi-channel access feature to provide EHT~\cite{ieee3}.
\begin{figure*}
	\centering
	\includegraphics[width=7in]{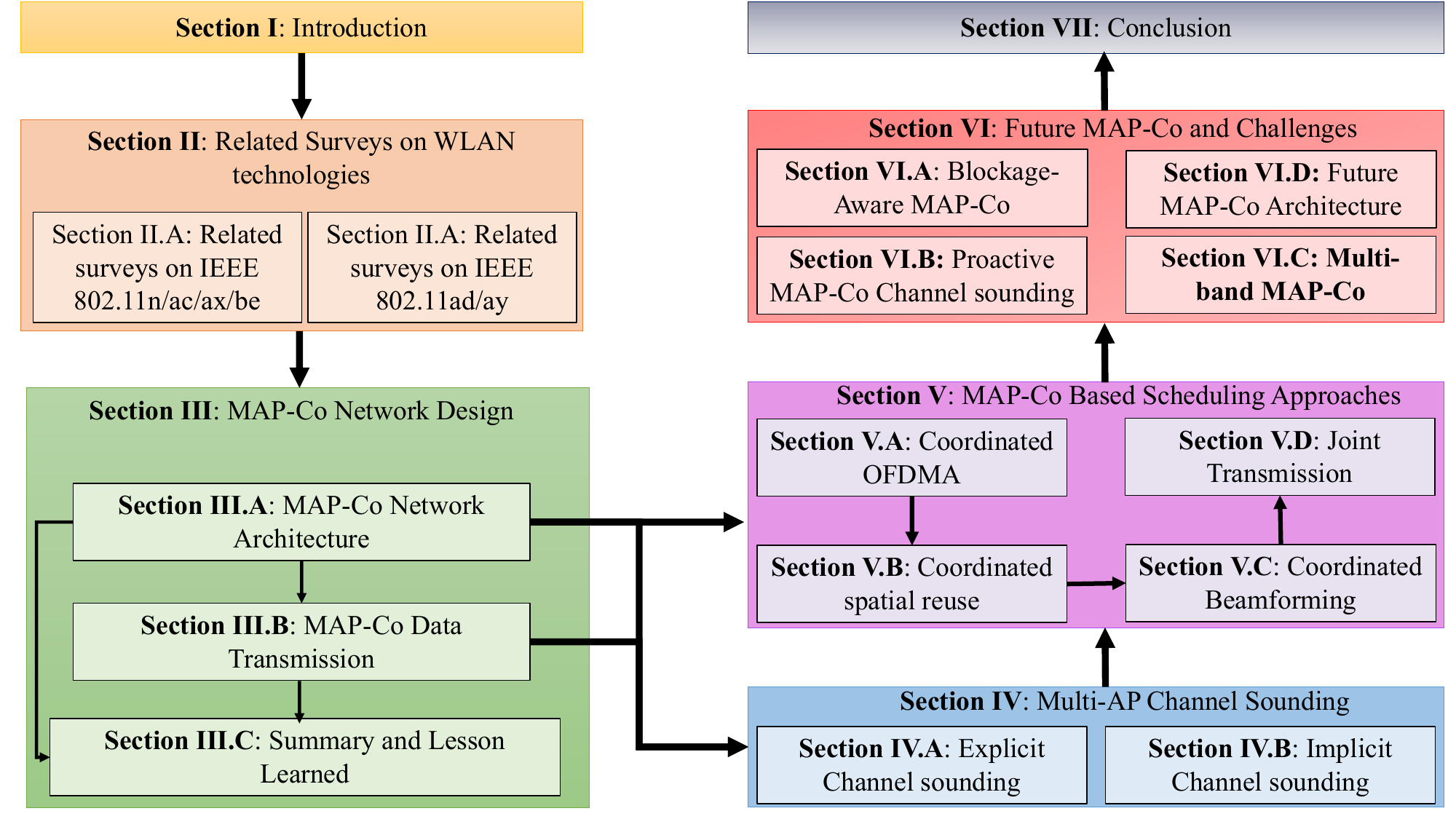}
	\caption{The overall structure of the paper}
	\label{structure}
	\vspace{-0.5cm}
\end{figure*}
The future of WLAN technology will also bring an unprecedented number of mobile users and access points (APs) that will need to support these users and share the massive network load. Additionally, the use of higher frequency bands with shorter transmission distances can also lead to a high density of APs~\cite{kobayashi,niyato2}. Despite advanced WLAN developments, such as multi-links and more channels per AP, users may still experience interference from neighboring APs, interrupted connections due to frequent handovers, and long waiting times to access an AP for transmission. These issues can negatively impact both the network throughput and latency.

Similar issues have also been encountered in cellular networks, leading to intercell interference, overloaded base stations, lower capacity, no coverage at cell boundaries, and channel collisions. Researchers have addressed these issues in cellular networks through coordinated multi-point transmission (CoMP), which uses smart antenna techniques like beamforming to improve coverage, and signal quality, and reduce interference. CoMP assigns appropriate modulation and coding schemes, allocates bandwidth, and manages transmit power levels among multi-cells to maximize performance and minimize interference~\cite{comp}. 
However, similar concept is not yet well developed for WLAN, and the concept of multi-AP coordination (MAP-Co) is being considered. CoMP and MAP-Co differ in several key aspects. CoMP operates on a larger scale within cellular networks, covering extensive geographical areas and often requires specialized infrastructure. MAP-Co functions within local area networks, typically encompassing smaller areas and can be implemented using standard Wi-Fi access points. Additionally, CoMP is usually managed and controlled by cellular network operators, while MAP-Co can be decentralized. Moreover, the PHY and MAC functions are different.
Therefore, the authors of \cite{kusano} proposed a new architecture to coordinate WiGig and sub-6GHz APs through MAP-Co. They compared a single-AP beamforming approach with a multi-AP approach in an indoor area measuring 10m x 20m with 8 WiGig APs and 24 devices. The results showed that coordinated WiGig LAN provided higher throughput and lower packet delay compared to non-coordinated single-AP transmission. MAP-Co has the potential to revolutionize IEEE WLAN standards and provide wireless networks with unparalleled capabilities. However, it is not widely studied within the scientific community and will not be included in the release of 802.11be and 802.11ay~\cite{kho,aysurvey}. We can expect that beyond IEEE 802.11be and 802.11ay/WiGig shall introduce MAP-Co.

 As there are many existing surveys on 802.11be and 802.11ay~\cite{aysurvey,deng}, yet most of them have reviewed overall PHY and MAC layer features of a single AP environment. Yet, there are no comprehensive surveys on MAP-Co for future WLAN and discussion on open issues for MAP-Co over next-generation WLAN. Fig.~\ref{feature-comparison} depicts the research gaps in the study of emerging WLANs. Hence, our focus is to firstly review the state-of-the-art on MAP-Co approaches that will be helpful for standards during the development of MAP-Co for future WLAN. In this article, we also present the future direction of MAP-Co in the emerging WLANs and discuss open issues to enable future MAP-Co.


\begin{table*}[!t]
	\renewcommand{\arraystretch}{1.3}
	\caption{A summary of the existing surveys related to emerging WLANs.}
	\label{related-surveys}
    \begin{center}
\begin{tabular}[c]{|c|m{5cm}|m{5cm}|m{3cm}|}

\hline
 Emerging WLANs &  \centering Title of the article & \centering Main scope & MAP-Co Study\\ \hline
\multicolumn{1}{|c|}{\multirow{6}{*}{IEEE 802.11be (Wi-Fi 7)}} & Wi-Fi Meets ML: A Survey on Improving \newline IEEE 802.11 Performance With Machine Learning~\cite{R9}             & Wi-Fi 7 PHY and MAC layers features where machine learning can be used for optimization. & There is no survey on MAP-Co. \\ \cline{2-4} 
\multicolumn{1}{|c|}{}                                        & IEEE 802.11be Wi-Fi 7: New Challenges and Opportunities~\cite{deng}                                          &Extensively review newly added PHY and MAC features of IEEE 802.11be, and discussed potential developments for beyond Wi-Fi 7. 
                 &  Briefly explained MAP-Co with respect to Wi-Fi 7 only.                 \\ \cline{2-4} 
\multicolumn{1}{|c|}{}                                        & Current Status and Directions of IEEE 802.11be, the Future Wi-Fi 7 \cite{kho}                      &The paper presented a brief tutorial on each novel feature of IEEE 802.11be.   & Authors didn't study individual MAP-Co methods and didn't address concerns for IEEE 802.11ay.                   \\ \cline{2-4} 
\multicolumn{1}{|c|}{}                                        & IEEE 802.11be Extremely High Throughput: The Next Generation of Wi-Fi Technology Beyond 802.11ax~\cite{lopez} & This article presents the key objectives, candidate features, enhancements, and coexistence issues of IEEE 802.11be with other 6GHz technology.              &The paper has provided tutorials on a few procedures of MAP-Co transmission without any discussion on issues. \\ \cline{2-4} 
\multicolumn{1}{|c|}{} & Overview and Performance Evaluation of Wi-Fi 7 \cite{R14} & This paper identified key PHY and MAC features of Wi-Fi 7, and enhanced quality of service with performance evaluation.        & There is no discussion on MAP-Co feature of Wi-Fi 7. \\ \cline{2-4} 
\multicolumn{1}{|c|}{} & Survey and Perspective on Extremely High Throughput (EHT) WLAN-IEEE 802.11be~\cite{R13} & Similar to the other articles, this article also briefly discussed key PHY and MAC features of Wi-Fi 7 and some issues without key details.        &There is a brief tutorial on MAP-Co transmission without deep new insights. \\ \hline
\multirow{5}{*}{IEEE 802.11ay/ad (WiGig)}                       & IEEE 802.11ay-Based mmWave WLANs: Design Challenges and Solutions \cite{aysurvey}                                &The article reviews MAC-related aspects of IEEE 802.11ay, including channel bonding, and so on. It identifies issues such as inter-AP coordination, high/low-frequency cooperation, and group formation. 
& There is a brief mention of Inter-AP cooperation as an open issue, no extensive survey.           \\ \cline{2-4} 
                                                              & Millimeter-Wave Fixed Wireless Access Using IEEE 802.11ay \cite{R19}  
                                                              &This magazine reviewed scheduling, beamforming and link maintenance of IEEE 802.11ay protocols to support mmWave fixed wireless access.   &There is no discussion on MAP-Co. \\\cline{2-4} 
                                                              & Wifi on steroids: 802.11ac and 802.11ad \cite{R17}                       & This article surveyed features of IEEE 802.11ac and 802.11ad such as MIMO, modulation and coding schemes, and beamforming. & The papers did not mention the need of MAP-Co in IEEE 802.11ay.                 \\ \cline{2-4} 
                                                              & Millimeter Wave Communication: A Comprehensive Survey~\cite{R18}                                                                                   &This article covers the use cases, PHY, MAC, and Network layers of mmWave WLAN using the IEEE 802.11ad protocol.             & The article did not provide any challenges of MAP-Co over mmWave networks.            \\ \cline{2-4} 
                                                              & IEEE 802.11ad: directional 60 GHz communication for multi-Gigabit-per-second Wi-Fi~\cite{R16}     & This invited paper provided a tutorial on use cases of IEEE 802.11ad, its PHY, MAC and network layer protocol, and beamforming concept.& The article did not present the idea and challenges of MAP-Co in WiGig.            \\ \cline{2-4} 
                                  & PHY/MAC Enhancements and QoS Mechanisms for Very High Throughput WLANs: A Survey~\cite{R4}                      & This survey examines how to enhance PHY/MAC protocols to improve quality of service satisfaction over the 60GHz band.  & This article also did not explain the scenario and challenge of MAP-Co over 802.11ay WLAN.  
                 \\ \cline{2-4}                             
                                                              \hline
\end{tabular}
 \end{center}
\end{table*}

The contributions of our work in this paper are as follows:
\begin{enumerate}
	\item 
	Through an extensive study of the existing surveys on WLANs, we are the first to point out the need for a comprehensive review of existing WLAN features concerning MAP-Co over future WLANs.
	
	\item 
	 This is also the first paper to provide a taxonomy and a comprehensive survey of MAP-Co architecture and existing MAP-Co features. In addition, we explain the shortcomings of each architecture and feature as a lesson learned. 
	\item 
	As the final contribution, we identify several future directions and several challenges for implementing future MAP-Co over the emerging WLAN. 
\end{enumerate}

The remainder of this paper is structured as follows, and can also be seen in Fig.~\ref{structure}: In Section~\ref{related-work}, we provide an overview of relevant surveys on emerging WLANs. In Section~\ref{overview-map-co}, we present a comprehensive survey on the architectures of MAP-Co network design, as well as a taxonomy of different processes for data transmission by MAP-Co. We also provide a summary of the lessons learned. Following that, in Section~\ref{channelsounding}, we provide an overview of the latest developments in multi-AP channel sounding, along with a summary of the lessons learned. In Section~\ref{maptransmission}, we discuss various MAP-Co scheduling approaches for data transmission, and after each discussion, we provide a summary of the lessons learned.
Section~\ref{futuresec} discusses several future directions and open research issues from the perspective of implementing MAP-Co over future WLANs. Finally, Section~\ref{Conclusion} concludes the paper. The acronyms used in this paper are summarized alphabetically in Table.~\ref{acronym}.



\section{Related Surveys}
\label{related-work}
This section provides related surveys on developed and emerging IEEE 802.11 WLAN standards. The subsequent subsection presents related surveys of 802.11n/ac/ax/be that operates over sub-6 GHz bands. The next subsection explains related surveys on mmWave WLAN standards that are 802.11ad/ay. At the end of both subsections, we present the gaps in existing literature and provide direction for further sections of this article.
\subsection{Related Surveys of IEEE 802.11n/ac/ax/be}
IEEE Standards and Wi-Fi Alliance are working on development of several WLAN technologies. The most used technology is the IEEE 802.11 family. IEEE 802.11n (also labeled as Wi-Fi 4) was one of the major leaps in WLAN technologies, which operated both on 2.4 GHz and 5 GHz. Several other WLAN standards followed 802.11n, operating on the same spectrum with new features. For instance, 802.11ac (Wi-Fi 5), and 802.11ax (Wi-Fi 6) are successors of 802.11n and are commercially available. There are already several surveys on 802.11n, 802.11ac, and 802.11ax. IEEE is also amending Wi-Fi 6 for the design of IEEE 802.11be (aka Wi-Fi 7). Hence, we present the related surveys on IEEE 802.11n/ac/ax/be in this section.\newline
Thomas \textit{et al.}~\cite{thomas} have presented a detailed survey on features of 802.11n, insights, and challenges of the PHY layer such as channel estimation, space-time block coding, and so on. Similarly, the authors in~\cite{xiao} have provided a comprehensive review of MAC layer enhancements for 802.11n. Further, Karamakar \textit{et al.}~\cite{karam} have presented a survey on enhanced PHY/MAC features of 802.11n/ac and studied the impact of new PHY features of 802.11n/ac on the upper layers of the network. There are also several surveys on PHY and MAC layer amendments of 802.11ac to design 802.11ax (Wi-Fi 6) ~\cite{R5,R6,R7,R8}.  For instance, the authors in~\cite{R5} have reviewed 802.11ax standardization activities of MAC protocols to better support QoS and explained the challenges of collaboration between cellular and 802.11ax. Bo \textit{et al.} \cite{R6} and Saloua et al. \cite{R7} surveyed different Orthogonal frequency-division multiple access (OFDMA) MAC protocols for allowing multi-user access and propose a new taxonomy to classify those methods. Evgeny \textit{et al.}~\cite{R8} have presented a comprehensive tutorial on IEEE 802.11ax features such as OFDMA-based random access, spatial frequency reuse, MU-MIMO, power saving, and so on. Most of these surveys focused on features under a single-AP transmission environment. Now, we discuss related surveys on the latest 802.11be protocol.

The first part of Table.~\ref{related-surveys} presents the related surveys of IEEE 802.11be/Wi-Fi 7. The authors in~\cite{R9} have provided a detailed survey on the application of machine learning for PHY and MAC features improvements such as channel access, link configuration, signal quality estimation, and so on. They also identified available tools and datasets while concluding with open research challenges of ML in Wi-Fi. However, the article lacks a thorough survey on new features of Wi-Fi 7.
\cite{lopez} is the first paper that summarized the objective, timelines, and listed candidate features of 802.11be. \cite{kho} is another tutorial article on the newly added PHY and MAC layer features of 802.11be. In this article, the authors listed multi-link, channel sounding, PHY and MAC format, enhanced OFDMA, and multi-AP cooperation. Regardless, the article failed to provide insights into the study of each feature and new open issues. In this regard, Cailian Deng \textit{et al.} \cite{deng} are the first to provide a more detailed review of MAC and PHY layer techniques mentioned by the task group. The discussed features are channelization and tone plan, multiple resource units (multi-RU) support, 4096-QAM, preamble designs, multiple link operations, MIMO enhancement, MAP-Co, enhanced link adaptation, and re-transmission protocols (e.g., hybrid automatic repeat request). The authors also provide a few insights into each feature with future directions and opportunities considering all features of 802.11be. However, such reviews considered all features which restricted them to provide a detailed review of all methods in each feature. After such wide reviews of all features, the authors are focusing on a thorough review of each feature. For instance, the authors of \cite{R13} gave a brief overview of features such as multi-band and multi-coordination concerning guaranteeing reliability, latency, and jitter. This paper focused on discussing a certain number of features with a particular objective. Hence, this article cannot be considered a wide review of a particular feature of 802.11be. \cite{R14} is one of the first articles to examine a specific feature of 802.11be. The authors of \cite{R14} briefly explained different multi-link operation modes and provided the performance evaluation. Likewise, there is a detailed survey on random access-based MAC protocols for MU-MIMO over 802.11ax by Liao \textit{et al.} \cite{R15}. They identified key requirements and research challenges of designing MU-MIMO. Such a detailed survey of a single feature will provide insightful contributions during the standardization of MU-MIMO. MAP-Co is a new concept that was never discussed in previous standards like MIMO, multi-RU, etc. Moreover, no detailed surveys are available on MAP-Co. Thus, the TGbe task group has decided not to include MAP-Co in the upcoming 802.11be standard which would be included MAP-Co as a part of the beyond 802.11be protocols with enough research. Hence, in this article, we provide an extensive review of every MAP-Co feature. Our comprehensive survey can provide a guideline for the future development of beyond 802.11be protocol. 
\subsection{Related Surveys of IEEE 802.11ad/ay}
In 2012, IEEE established another task group to focus on designing 802.11 WLAN standards for accessing unlicensed mmWave such as 60 GHz spectrum to guarantee Gbps throughput. In this vein, IEEE introduces the 802.11ad standard intending to design the operation of Wi-Fi that can address the several challenges of accessing mmWave bands. However, there are a few good surveys that provide an extensive survey of IEEE 802.11ad features. For instance, \cite{R4} is the first paper that provided a succinct study of the PHY and MAC features of 802.11ad along with 802.11ac. They explained PHY and MAC layer format, beamforming training, different aggregation approaches, channel access mechanisms, and QoS mechanisms. However, the article did not provide any future direction and open challenges for future 802.11ad protocols. The survey in \cite{R16} provided a detailed tutorial on 802.11ad wherein the authors enumerated some design challenges for 802.11ad such as directional transmission using beamforming, new PHY and MAC layer format, and channel access methods.  Verma \textit{et al.} \cite{R17} provided an overview of 802.11ac/ad features and hardware challenges such as hardware complexity, and semiconductor cost. Similarly, \cite{R18} has surveyed use cases, PHY, MAC, and Network layers of 802.11ad protocol. In PHY layer, the authors have focused on MIMO, channel model, precoding and so on. In the MAC layer, the paper reviewed MAC layer protocol in different types of networks such as ad-hoc networks, mesh networks, etc. However, these articles failed to provide future issues and directions to address the challenges of designed 802.11ad protocols. 802.11 standards have proposed amendments on 802.11ac to enhance the MAC and PHY layer protocols for supporting next-generation applications such as AR/VR, data center connectivity, vehicle-to-vehicle connection, etc. The new mmWave WLAN standard is named 802.11ay. In this vein, \cite{R19} is the first paper to list newly incorporated features on 802.11ad to design 802.11ay. This review article focused on beamforming enhancements to support new applications, and different resource scheduling approaches. However, the authors did not provide any open issues that are not mentioned by the standard.

Finally, Zhou \textit{et al.} \cite{aysurvey} conducted a broad review on the MAC-related issues of 802.11ay, cross-layer issues between MAC and PHY while reviewing the challenges of 802.11ad MAC protocols that lead to the design of 802.11ay MAC. The authors explained channel bonding/aggregation, channel access and allocation using MIMO over multiple channels, spatial sharing, interference mitigation approach, beamforming training and tracking, and single and MU-MIMO. Additionally, they pointed out open research issues and future work. They have mentioned that inter-AP coordination/MAP-Co will be one of the significant future directions for IEEE 802.11ay owing to several advantages and the ability to solve the existing issues. The MAP-Co can reduce the handover delay, improve failed link recovery, much higher throughput using joint transmission (JTX), inter-cell interference (ICI) avoidance, and so on. However, the authors did not explain any technical details and challenges of implementing MAP-Co in 802.11ay. Based on the authors’ knowledge and the summary presented in the third column of Table~\ref{related-surveys}, it seems that no review article has thoroughly examined the future challenges and implementation of MAP-Co for 802.11ay. That being said, we also aim to fill this gap by providing a comprehensive review of MAP-Co in the context of the future 802.11ay protocol. To get an overview of the existing surveys on emerging 802.11 standards, please refer to Table~\ref{related-surveys}. Therefore, it is evident that there is a need for an extensive survey on MAP-Co among emerging WLANs.

The ongoing research on MAP-Co focuses on multiple aspects, including architecture design, multi-AP channel sounding, multi-AP coordinated resource allocation, and JTX by multi-APs~\cite{lopez}. The literature has identified two prominent architecture types: centralized and distributed architecture~\cite{ryu,perez}. Therefore, the next section outlines the existing architecture of MAP-Co. Once the network architecture is defined, it is essential to explain how multiple APs access physical resources to transmit data based on channel conditions. Therefore, we first provide an extensive literature review of different channel sounding approaches to acquire information about channel conditions. Several works have been published on multi-AP channel sounding~\cite{park,ryu}. After acquiring channel state information (CSI), network resources must be managed among multiple APs to avoid interference and provide good Signal-to-Interference-plus-Noise Ratio (SINR). In this vein, researchers have studied coordinated OFDMA (C-OFDMA), coordinated spatial reuse (CSR), coordinated beamforming (CBF), and JTX~\cite{cofdma6,CSR1,cbfwlan,deng}. Therefore, in section~\ref{maptransmission}, we provide the state-of-the-art of C-OFDMA, CSR, CBF, and JTX. We also present lessons learned from discussions of multi-AP channel sounding and MAP-Co data transmission. Finally, in the next section, we present future directions and challenges of MAP-Co over emerging WLANs.

\begin{figure}
	\centering
	\includegraphics[width=3.5in]{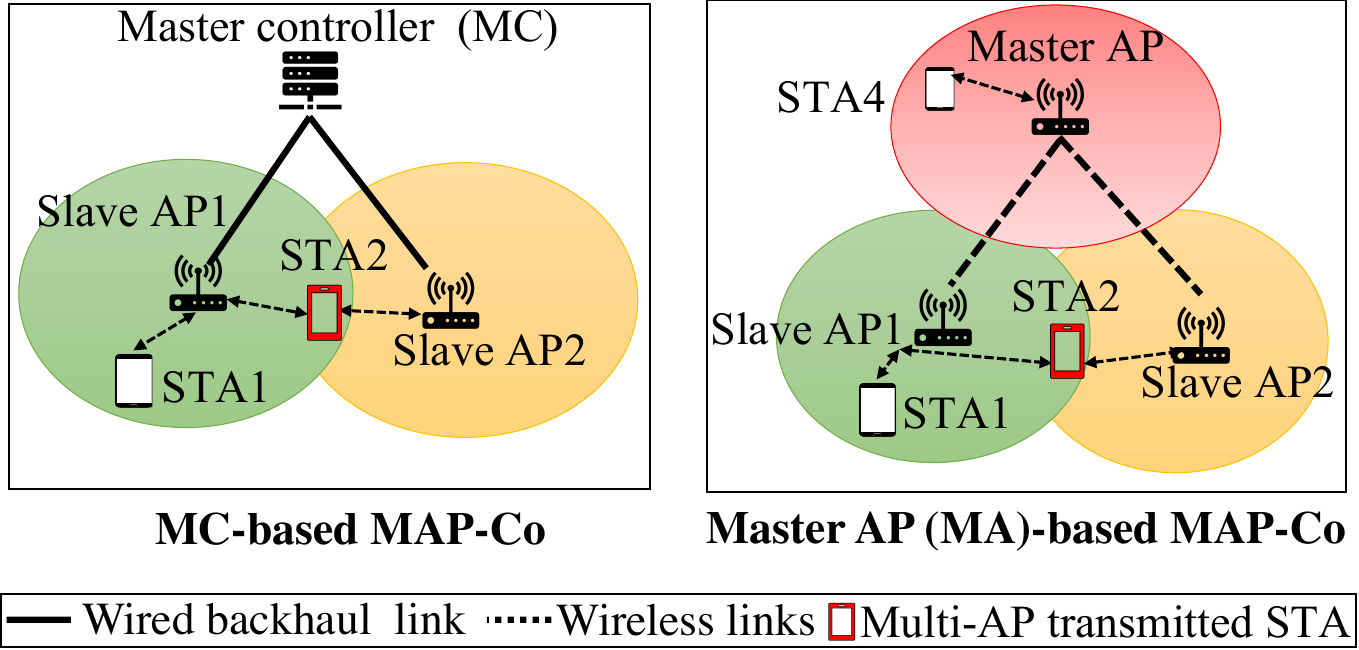}
	\caption{Network architectures of MAP-Co: MC-based MAP-Co and MA-based MAP-Co.}
	\label{arch}
	\vspace{-0.5cm}
\end{figure}
\section{MAP-Co Network Design}
\label{overview-map-co}
The 802.11be technology introduces a new concept called MAP-Co, which is crucial for coordinating neighboring APs, allowing for multiple APs to transmit data without interference and collisions. Therefore, it is essential to study the network design of this upcoming technology that can revolutionize WLAN. Network design involves two steps - defining the network architecture and then establishing data transmission protocols. In this section, we will focus on exploring the MAP-Co network design architecture, which serves as the foundation for studying various MAP-Co data transmission protocols in later sections. The objective is to comprehend how the coordinator and coordinating APs interact to determine the most efficient way to enable multi-AP data transmission. Therefore, understanding the architecture is crucial before investigating various MAP-Co-based data transmission processes, which we will outline briefly in Section \ref{protocols}. Finally, the conclusion will summarize the lessons learned from reviewing MAP-Co architectures and the taxonomy of MAP-Co-based data transmission.
\subsection{MAP-Co Network Architecture}
A typical multi-AP network scenario consists of several neighboring overlapping APs and distributed stations (STAs) under each AP. For instance, a large factory or shopping mall has several APs to manage connections for static and mobile users. The coordination between multi-APs can improve reliability, reduce latency, increase manageability, increase throughput at different signal-to-noise-ratio (SNR), and reduce power consumption \cite{deng}. For instance, multi-AP coordinated transmission and reception increase throughput, especially in the case of cell edge users having high path loss and interference from neighboring APs. However, the traditional neighboring APs can be developed by the same or different original equipment manufacturers . However, the diverse APs can have interoperability issues and previous standards had no general protocol for leveraging coordination between APs. Therefore, 802.11be introduces new components, methods, MAC, and PHY layer protocols to enable MAP-Co. In MAP-Co, there is a master controller as a multi-AP coordinator and neighboring APs that can called as slave APs~\cite{ryu}. Slave APs are connected with a master controller through Ethernet or to an master AP within its range. STAs are connected to slave APs and the master controller/master AP cannot hear STAs directly unless STAs are directly connected to it. However, a master controller/master AP can have information of STAs through their connected slave APs. Hence, a master controller/master AP, slave APs, and their associated STAs are called multi-AP candidate sets, and their information is maintained by the master controller/master AP. The master controller/master AP can also be singly called as coordinator. Thus, there can be two ways of interconnecting the coordinator, slave APs and STAs. These two approaches define the architecture of MAP-Co and are named as master controller-based (MC-based) and master AP (MA)-based MAP-Co\cite{ryu,perez}. MC-based is a centralized whereas MA-based is a semi-distributed architecture. Both architectures are depicted in the Fig.~\ref{arch}. 

In MC-based, the master controller can be a local server, where software-defined networking functions are defined and slave APs are connected to the master AP through high capacity, low latency fiber or wireless backhaul links, as shown in Fig.~\ref{arch}. The master controller has a full view of all its connected slave APs and their associated STAs. The backhaul links are used to shared information by slave APs such as resources utilized, managing carrier sense multiple access/collision avoidance (CSMA/CA) states, CSI and so on. The master controller sends control information or request for CSI updates to slave APs. Slave APs can report CSI and schedule transmission based on received control signals from MC. 
The authors of \cite{lzhang} proposed a new system architecture that uses MC-based or centralized architecture to enable MAP-Co, which can improve network efficiency, minimize contention, and enhance reliability in multi-AP 802.11be networks. The proposed system architecture includes a Centralized AP Controller (APC) that acts as a master controller and assigns primary channels to APs in a fair manner. This approach eliminates contention among APs for wide-band transmission opportunities (TXOP) and simplifies the design of ACK. 
The APC performs radio resource management by monitoring traffic, capacity, and reliability of operating APs. It carries out periodic reconfiguration of the 802.11 networks to achieve optimal efficiency by performing tasks such as radio resource monitoring and dynamic channel assignment. Each AP competes for TXOP on the assigned primary channel, which is further used for communication with its associated STAs using UL/DL OFDMA. 
Through their study, the authors conclude that the proposed architecture is essential for optimizing channel access by multiple APs.

In the MA-based approach, an AP is chosen as a coordinator which can be connected to multiple slave APs and called as a master AP~\cite{cai, woo}. The architecture of MA-based MAP-Co is shown in Fig.~\ref{arch}. As depicted in Fig.~\ref{arch}, there are three connected basic service sets (BSS), where AP1, AP2 and AP3 are access points in each BSS. STA1 and STA2 are associated with AP1, STA2 can be also associated with AP2, and STA3 is associated with AP3. AP3 is the MA whereas AP2 and AP3 are slave APs. STA 2 is at the edge of a cell and under the range of AP1 and AP2. STA2 can get the benefits of multi-AP transmission whereas other STAs will have single AP transmission, as presented in Fig.~\ref{arch}. One way to select a master AP is by considering transmission opportunities. The process starts with each AP fairly competing for the channel using CSMA/CA. If an AP wins the TXOP over the channel, it becomes the master AP for that coordinated transmission. The master AP then becomes the scheduling leader and shares its allocated resources with other APs, which are referred to as slave APs~\cite{cai,cofdma6,tanaka}. Another approach to selecting a master AP is to use the round-robin method. This method ensures fairness in multi-AP transmission while minimizing energy consumption \cite{woo}. In some cases, the role of master and slave APs can change depending on the training time for executing the federated learning algorithm. In such cases, multi-AP coordination is used to perform federated learning and share the model weights. The primary AP is chosen based on the training time at each AP, the workload at each AP, and the energy consumption associated with devices. The AP with the longest training time is chosen as the master, while slave APs are chosen based on the shortest training time, as the federated learning algorithm operates only on slave APs~\cite{woo}.
 Based on the network conditions, the master AP then selects the slave APs for coordinated transmission using a selection algorithm. The master AP sends a trigger frame to each slave AP, indicating that they have been selected as the slave AP and that they should transmit data together within a specific time period. After receiving the trigger frame from the master AP, the slave AP prepares for data transmission with the master AP. Both the master and slave APs start data transmission simultaneously after a set time. 
Hence, the MA should perform extra MAC and PHY layer functions of coordination during each data transmission. Slave APs can also work as traditional single AP transmission. Both of the architecture has advantages and drawbacks. Hence, at the end of this section, we summarize the architectures with advantages and drawbacks as lesson learned from study of each architecture. Moreover, once the network architecture is defined, the next step is to determine the process of network operations and data transmission by multiple APs. Therefore, in the next section, we will elaborate on the taxonomy of MAP-Co features required for efficient MAP-Co operations. We will also explain each feature in relation to different architectures in sections V and VI.

\subsection{MAP-Co Data Transmission}
\label{protocols}
In wireless communication with a single AP, the first step is to gather CSI from various STAs to improve transmission. With this information, we can allocate resources, control transmission power, and optimize beamforming processes for transmissions in specific directions. When multiple APs are coordinated, the same wireless communication processes need to be coordinated among them to provide high performance for MAP-Co. Therefore, the channel sounding information by multiple APs to all of their STAs needs to be shared among each AP, which is one of the primary steps of MAP-Co. The multi-AP channel sounding process depends on the MAP-Co architecture, due to the different connections between a master and its slaves, and the starting of channel sounding from the master. There are two ways of channel sounding: Explicit and Implicit. Therefore, we provide extensive review of explicit and implicit multi-AP channel sounding in Section IV. Moreover, Section IV will delve into the MAP-Co channel sounding methods for different types of MAP-Co architecture discussed in Section III.A. We will also discuss the benefits and drawbacks of each sounding method for different architectures. The relationship between different sounding methods and architecture is presented in detail for each sounding method mentioned in Section IV. This study can help determine the appropriate channel sounding method for each type of MAP-Co architecture.
Thereafter, the coordinated resources scheduling by each AP are the next step for data transmission. Without coordination, resource allocation can lead to co-channel interference such as allocating neighboring resource blocks or increasing waiting time for priority services. Therefore, coordinated resource allocation specifically C-OFDMA is discussed after multi-AP channel sounding in Section V.A. Only coordinated resource allocation cannot avoid the interference among multiple APs and poor SINR at users. Therefore, the transmission power also needs to be controlled in a coordinated way for CSR. For instance, some STAs of APs can have poor SNR due to low receiving power from their APs. Due to this reason, we explain CSR after C-OFDMA. Beamforming is also used for spatial reuse by keeping the same transmission power. In the case of beamforming, the side and back lobes of beams can interfere with the other APs' transmission or transmission at other APs. Therefore, the beamforming approach needs to be coordinated to nullify or cancel interference and simultaneously to the transmission. Hence, we discuss CBF after CSR. Even after optimizing multi-AP channel sounding, C-OFDMA, CSR, and CBF, the AP cannot guarantee a successful connection and high network performance to all users. For instance, the edge users cannot get good SINR from the connected AP and can have unsuccessful transmission until they hand over to the new AP. MAP-Co can solve this issue by allowing STAs to connect to multiple APs simultaneously. So, users can always get the data from an AP without handover even if it is the edge of another APs. This feature of MAP-Co is called JTX and is explained at last of Section V. 
To achieve overall improvement of data transmission by each AP and user, we discuss all the MAP-Co features in section IV and V. These features can take the evolution of WLAN to new heights by providing high performance anytime and anywhere over a large area network.
\subsection{Summary and Lesson learned}
The architecture discussion highlights that MC-based approach is centralized, while MA-based approach is semi-distributed. MC offers greater computational power and higher bandwidth compared to MA, as it operates on local servers and wired backbone networks, whereas MA uses an AP as its controller and wireless connection. However, implementing MC requires additional infrastructure, making it expensive, while MA-based approach utilizes slave APs for coordination, making it cost-effective. MC-based architecture provides better control of resources and synchronization between multiple slave APs. In contrast, MA-based systems can be more complex to manage and synchronize due to the distributed nature of resources. Additionally, MA-based architecture is more scalable than MC systems, as the load can be distributed among multiple slave APs rather than being congested at a centralized server. In the future, WLAN will have a high density of various small cells like THz, mmWave, and sub-6GHz. However, in these high-density areas, it can be challenging to coordinate with all APs within their range, causing significant overhead. To tackle this issue, the authors in \cite{kihira} introduced a RARL-based approach that trains both a central agent (for coordinated APs) and an adversarial AP (to disturb coordinated APs). This method leverages the history of frame losses from coordinated APs to enhance competition between the central agent and the adversarial AP. The simulation results demonstrated that the proposed method effectively avoids uncoordinated interference and improves throughput compared to not considering uncoordinated APs. This RARL-based approach can help avoid including all neighboring APs to reduce the scale at MC-based architecture and mitigate interference from those APs.
Additionally, MC-based systems offer better security as all slave APs can be controlled more easily, while MA-based systems may have less security due to their distributed nature.
In this section, we also discussed different data transmission processes of a MAP-Co system and the relationship between each process. The first process is multi-AP channel sounding~\cite{ieee4,sch}. Hence, we first provide a comprehensive survey of multi-AP channel sounding, different approaches, advantages, and drawbacks in the next section. Moreover, we have also observed from \cite{lzhang,cai}, that the process of network operations for multi-APs varies with each architecture. Therefore, we will also explain different channel sounding methods in relation to different architectures in section~\ref{channelsounding}. In section V, we will explain MAP-Co data transmission scheduling approaches that are C-OFDMA, CSR, CBF, and JTX. We will also analyze each method with respect to two architectures explained in section III.A.

 \begin{figure*}
	\centering
	\includegraphics[width=7.3in]{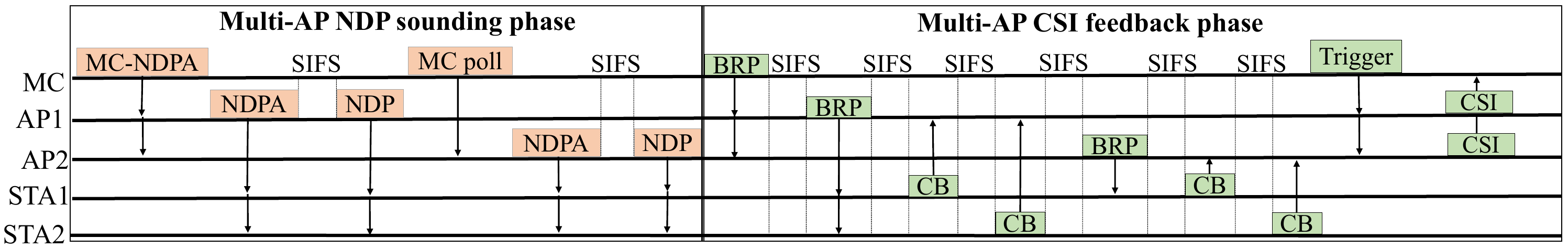}
	\caption{Multi-AP MC-based sequential explicit channel sounding.}
	\label{MC-explicit}
	\vspace{-0.5cm}
\end{figure*}
 \section{Multi-AP Channel Sounding}
 \label{channelsounding}
Channel sounding is a well-known process of surveying radio frequency channel characteristics and beam training over WLAN~\cite{ieee6,ieee7,ieee8}. The channel quality can be significantly affected by obstacles, interference, propagation paths, diffraction loss, reflection loss, and so on. The channel sounding helps to adapt the data transmission according to the current propagation environment. For instance, the case of signal loss below a threshold can indicate to increase in the transmission power or a change in modulation and coding values. Hence, channel sounding is a key process for efficient data transmission by accurately estimating CSI. The CSI feedback acquisition is the process of channel sounding. The collected feedback can also be used to decide a weight and steering matrix for the optimization of beamforming also. In a single AP WLAN, the sounding procedure is carried out within a BSS~\cite{ieee8,kho} that is between an AP and STAs. The first step is to send a request for CSI feedback to STAs and STAs respond with a null signal to AP. The AP evaluates the channel quality based on a received signal. Another approach is that STAs estimate channel states based on the received request signal and feedback CSI to APs as a response. Hence, there are two existing CSI gathering schemes in the literature: Explicit and Implicit channel sounding. Channel sounding approach is also known as CSI feedback acquisition in other WLANs such as 802.11ay. We refer CSI estimation, channel quality estimation, etc as channel sounding in the rest of the article. However, Multi-AP transmission approaches such as CSR, CBF, JTX, and so on, require CSI between STAs and different APs beforehand. The process of acquiring CSI information between different STAs and multiple APs is called as MAP-Co channel sounding. The existing channel sounding process developed for a single-AP environment is not suitable for MAP-Co because the process involves signaling from an AP to its only associated STAs. Moreover, the architecture of MAP-Co is different, which consists of master AP and slave APs whereas there is no coordination in a single AP. 

  Therefore, the 802.11be WLAN discussed new processes for the MAP-Co channel sounding. Similar to single-AP channel sounding, MAP-Co channel sounding should be two types:  Explicit MAP-Co channel sounding and implicit MAP-Co channel sounding. This section presents comprehensive surveys on such two types of channel sounding concerning MAP-Co while showing how the single-AP sounding method is improved to enable MAP-Co channel sounding in the next subsections. We also summarize and provide lessons learned from each discussed multi-AP channel-sounding approach.
 
 \begin{figure*}
	\centering
	\includegraphics[width=7.2in]{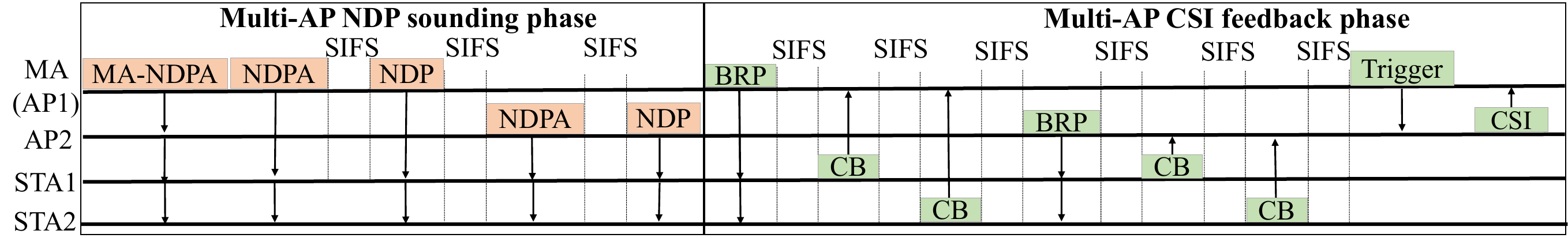}
	\caption{Multi-AP MA-based sequential explicit channel sounding.}
	\label{MAP-explicit}
	\vspace{-0.5cm}
\end{figure*}
 \subsection{Explicit channel sounding}
 Explicit channel sounding requires the receiver to estimate CSI and periodically feedback to a transmitter, as mentioned before when STAs estimate CSI to report back to AP for downlink (DL) transmission~\cite{ieee8}. 
 In the 802.11be single-AP system, the process of explicit channel sounding begins when the AP sends a null data packet announcement (NDPA) to one or more STAs. This is followed by a short interframe space (SIFS) and then the transmission of a null data packet (NDP) frame, which only contains the physical header. The STAs that received the NDPA frame wait for the NDP frame, and the first STA mentioned in the NDPA frame sends compressed beamforming (CB) frame after another SIFS. The CB frame includes a feedback matrix that informs the AP about beam steering for future data transmission. The other STAs wait for a beamforming report poll (BRP) frame from the AP to send their CB feedback. This process is repeated for each STA mentioned in the NDPA frame. Once the AP collects CB from all the STAs, it can transmit data to multiple users or a single user using the received CB. In the case of UL MU-MIMO or UL OFDMA, the CB from multiple users can be transmitted together, reducing the CSI feedback overhead in a single-AP system. However, in the case of MAP-Co, all participating APs must be aware of the CSI at each STA in their range to avoid issues like ICI and resource utilization~\cite{park,jyu,ayfeature}. Therefore, the 802.11be introduces a multi-AP explicit channel sounding MAC process.

     There are two architectures in MAP-Co transmission: MC-based and MA-based. To classify the process of explicit MAP-Co channel sounding, we differentiate between these two architectures. The channel sounding process differs slightly in each MAP-Co architecture. In the MC-based architecture, the master controller (MC) sends a trigger frame or NDPA frame (called MC-NDPA in this article) to each AP in its control (Slave APs) to initiate sounding, as shown in Fig.~\ref{MC-explicit}.

The MC-based architecture involves slave APs transmitting NDP frames either sequentially or jointly after receiving MC-NDPA from the MC. This results in two types of multi-AP explicit sounding: sequential and joint MC-based. In sequential multi-AP explicit channel sounding, a slave AP broadcasts an NDPA frame to STAs, followed by an NDP frame to all STAs after SIFS. Once the transmission of NDP frames is complete, MC sends the MC poll frame to the next slave AP in the sequence mentioned in the MC-NDPA transmit. After receiving the MAP poll frame, the slave AP broadcasts NDPA frames to STAs and transmits NDP frames after SIFS. This process is repeated by other slave APs in sequential multi-AP explicit channel sounding. The process of joint MC-based explicit channel sounding involves slave APs broadcasting NDPA and NDP frames together. This phase, in both types of sequential and JTX, is called the multi-AP NDP sounding phase, and it is depicted in Fig.~\ref{MC-explicit}.

Following the initial phase, the multi-AP CSI feedback phase takes place. In MC-based systems, MC sends out the BRP trigger frame to slave APs after the NDP sounding phase. With sequential multi-AP explicit channel sounding, the first slave AP on the list of MC-NDPA frames forwards a BRP frame to its STAs. The STAs respond to the BRP frames with CB frames, each with information about the channel quality and beamforming parameters, sequentially with a SIFS gap. Once the CB frames from the first slave AP are complete, the next slave AP performs the same process in sequence, followed by the other slave APs in a similar fashion. In joint multi-AP explicit channel sounding, the slave APs send the BRP frame to all STAs simultaneously. After the SIFS interval of receiving BRP frames, the STAs transmit CB frames together to all slave APs.
 Thereafter, MC sends a trigger frame to collect the CSI information of all STAs from each slave AP. In sequential, each slave AP transmits the CSI feedbacks individually whereas slave APs report CSI information to MC simultaneously in joint multi-AP explicit channel sounding. 
 
In the MA-based architecture, the MA sends an NDPA trigger frame to slave APs and associated STAs for channel sounding. The sequential MA-based channel sounding involves the MA and slave APs performing the NDP sounding phase after the NDPA. First, the MA transmits NDPA and NDP frames to slave APs and STAs with a SIFS interval. Then, slave APs send NDPA and NDP frames sequentially. After the NDP sounding phase, STAs of MA report CB sequentially with a SIFS interval, using the BRP frame to instruct them. Each slave AP transmits a BRP frame to each STA after receiving it from MA. This frame indicates slave APs to request CB to their associated STAs sequentially, with each STA reporting CB information in the CSI feedback phase. After receiving CB frames from STAs, MA sends a trigger frame to slave APs, and slave APs report CSI to MA sequentially. In joint explicit MA-based channel sounding, NDPA, NDP, and CB frames are transmitted by MA, slave APs, and STAs together during their turn. MA requests slave APs through a trigger frame to report collected CB from different STAs, and all slave APs transmit CB reports to MA jointly with a SIFS interval.
 \begin{table*}[]
\renewcommand{\arraystretch}{1.3}
	\caption{Taxonomy of channel sounding method in MAP-Co, along with their pros and cons.}
	\label{sounding-comparision-table}
\begin{tabular}{|m{2.5cm}|m{2.5cm}|m{2cm}|m{4cm}|m{5cm}|}
\hline
Channel Sounding                                & Architecture type                 & Sequential/Joint & Advantages                                                           & Drawbacks                                                                                                                       \\ \hline
\multicolumn{1}{|l|}{\multirow{4}{*}{Explicit}} & \multirow{2}{*}{MC-based MAP-Co}  & Sequential       & No network congestion, lower computation overhead,                   & Longer waiting time/latency, inefficient resource utilization, Not designed for extreme high frequency WLAN such as 802.11ay.                                                            \\ \cline{3-5} 
\multicolumn{1}{|c|}{}                          &                                   & Joint            & Lower waiting time and can handle computation overhead               & Network congestion, not suitable for applications having massive STAs/AP, Not designed for extreme high frequency WLAN such as 802.11ay.                                                         \\ \cline{2-5} 
\multicolumn{1}{|c|}{}                          & \multirow{2}{*}{MAP-based MAP-Co} & Sequential       & No network congestion, lower latency than sequential MC-based MAP-Co & Longer waiting time, higher computation overhead than sequential MC-based, inefficient resource utilization, Not designed for extreme high frequency WLAN such as 802.11ay .                    \\ \cline{3-5} 
\multicolumn{1}{|c|}{}                          &                                   & Joint            & Lower waiting time and efficient resource utilization                & Higher network congestion compared to joint MC-based MAP-Co, energy inefficient, higher computation and communication overhead, Not designed for extreme high frequency WLAN such as 802.11ay. \\ \hline
\multirow{4}{*}{Implicit}                       & \multirow{2}{*}{MC-based MAP-Co}  & Sequential       &    Lower calibration processing overhead, centralized approach, and lower UL network congestion, and No computational overhead issue.                                                                 & Longer waiting time and delay, and inefficient resource utilization.                                                                                                                                 \\ \cline{3-5} 
                                                &                                   & Joint            &    Centralized, efficient resource utilization, multi-channel sounding, No computational overhead for calibration, and lower delay.                                                                  &  Network congestion, and higher chance of calibration errors.                                                                                                                               \\ \cline{2-5} 
                                                & \multirow{2}{*}{MAP-based MAP-CO} & Sequential       & Lower communication overhead compared to MC-based, lower transmission delay, and no UL network congestion.             &  Higher calibration computation overhead at MAP, longer processing delay, and inefficient network resource utilization.                                                             \\ \cline{3-5} 
                                                &                                   & Joint            & Efficient resource utilization, lower waiting delay, and multi-channel sounding       & Highest calibration calculation processing overhead compared to all implicit methods, network congestion, higher calibration errors chances, and energy inefficient.                                                                                                                                \\ \hline
                                                                       
\end{tabular}
\end{table*}

\begin{table*}[!t]
	\renewcommand{\arraystretch}{1.3}
	\caption{State-of-the-art of different methods to reduce the explicit channel sounding overhead.}
	\label{explicit-overhead}
\begin{tabular}{|m{3cm}|m{3.5cm}|m{4cm}|m{4cm}|m{1.5cm}|}
\hline
Scheme                    & Method                       & Advantage                                                   & Disadvantage                                                                              &MAP-Co suitability  \\ \hline
\multirow{4}{*}{Enhanced explicit} & $\phi$ Only feedback \cite{E1}               & Existing in 802.11ah with minor change in MAC protocols     & This method only supports one data stream and does not reduce overhead significantly. & $\times$ \\ \cline{2-5} 
                          & Time domain channel feedback \cite{E2}    & Existing in 802.11ad/ay and overhead is lower               & Need extra signalling to identify transpose and extra matrix.                             & \checkmark \\ \cline{2-5} 
                          & Differential Given rotation \cite{E3}      & Reducing overhead significantly in 802.11ax/be and 802.11ay & Need additional processing and can experience propagation errors.                          & $\times$ \\ \cline{2-5} 
                          & Variable angle optimization  \cite{E4,E5}     & Only designed for 802.11ax                                  & Also requires additional processing and signaling.                                        & \checkmark \\ \hline
\multirow{3}{*}{New Schemes}   & Multiple component feedback \cite{E4}       & Reduce feedback overhead                                    & Requires to redesign MAC to change feedback sizes and indication of intervals.            &\checkmark  \\ \cline{2-5} 
                         & Codebook based feedback   \cite{E6}                       & Well studied and reduced feedback overheadL & Require additional processing power and new design.                 & \checkmark  \\ \cline{2-5} 
                         & Deep learning     \cite{E7,E8}             & Predicting CSI based on previous patterns can significantly reduce feedback overhead, making it a great fit for MAP-Co. & To store previously reported CSI, additional processing and storage are required. It may not be feasible for an MA-based architecture.                                                         &\checkmark  \\ \hline
\end{tabular}
\end{table*}
 
 \subsubsection{Summary and Lesson Learned}
In the beginning, we categorize the explicit channel sounding for MAP-Co into two types: MC and MA-based. We then proceed to describe the process for each type of explicit channel sounding in MC and MA-based MAP-Co.
In both architectures, we talked about two types of channel sounding: sequential and joint explicit channel sounding. Sequential involves each AP and STA taking turns transmitting channel sounding frames, while joint sounding sends all frames from slave APs and STAs at once.

Based on the discussion, we can conclude that using a sequential sounding approach for channel sounding requires more frame exchanges. As a result, this method can significantly increase communication and computation overhead for MC and MA. MC-based architecture can handle this overhead better because it only communicates with slave APs and has higher processing power. However, MA-based architecture may struggle with high overhead because it has to process all slave APs and STAs. Sequential explicit channel sounding has the benefit of not congesting network resources, making it the best option for MC-based architecture. However, it has a longer waiting delay due to waiting for other slave APs and STAs to finish their sounding process, resulting in higher latency. This method is not suitable for real-time or ultra-low latency services and does not meet the requirements of higher bandwidth for future applications like augmented reality and massive IoT. 

On the other hand, joint explicit channel sounding allows for joint exchange of frames by MC, MA, slave APs, and STAs, making it a better approach for proper resource utilization of OFDMA-based resource allocation. However, it can highly congest the WLAN in the case of massive IoT applications per AP, leading to unsuccessful transmission and re-transmission, which increases communication overheads, energy consumption, and delays actual data transmission~\cite{verma2}. Additionally, JTX is not applicable for energy and delay constrained STAs, and computation and communication overhead in explicit channel sounding for MAP-Co is an issue at a larger scale compared to a single AP explicit channel sounding. In Table~\ref{sounding-comparision-table}, the benefits and limitations of each explicit channel sounding type in MAP-Co are outlined. It is clear from the analysis that the explicit channel sounding method incurs a significant overhead due to the increased number of frame exchanges required.

Different methods have been proposed to decrease signal overhead in a single AP environment in 802.11ax/be. Table~\ref{explicit-overhead} summarizes the state-of-the-art overhead reduction methods for explicit channel sounding designed for 802.11ax/be and determines the best solution for MAP-Co WLAN environment. These methods have been well-researched in the literature~\cite{E1,E2,E3,E4,E5,E6}, so we won't go into detail about each one. Based on the information in Table~\ref{explicit-overhead}, we'll explain what we learned from each solution regarding its suitability for MAP-Co channel sounding. 

The time parameter-based solution proposed in~\cite{E1} is suitable for MAP-Co because it reduces the amount of information needed to encapsulate channel information such as multipath locations or amplitude by transforming frequency domain channel observation into the time domain. However, the approach used in \cite{E2} is not well-suited for MAP-Co channel sounding because it's designed for single data streams, which won't be the case in future WLANs. The solution presented in \cite{E3} can't solve issues in MAP-Co due to the need for additional processing and signaling, which could lead to errors in estimation. Similarly, the solution in \cite{E4} requires additional processing and signaling, which may not be good for MA-based MAP-Co channel sounding. However, it can be applied for MC-based explicit channel sounding since higher resources are available. 

The proposed method in \cite{E5} can be used for MC and sequential MC-based MAP-Co systems since it dynamically sets the size of CSI feedback. However, reducing feedback size alone won't be enough for joint and MA-based explicit channel sounding. Another idea to decrease overhead based on feedback is using codebook~\cite{E6}. The feedback overhead can be decreased by optimizing the size of the codebook, which sounds promising for MAP-Co. However, it requires additional processing power and a new design for the MAP-Co system. Recently, machine learning-based approaches have been proposed for CSI estimation in different systems, which significantly reduce signaling overhead~\cite{E7,E8}. For example, \cite{E7} proposed a fully connected deep neural network for CSI estimation in cellular networks, but such an approach requires a lot of processing power that may not apply to MA-based architecture.
 \begin{figure*}
	\centering
	\includegraphics[width=7.2in]{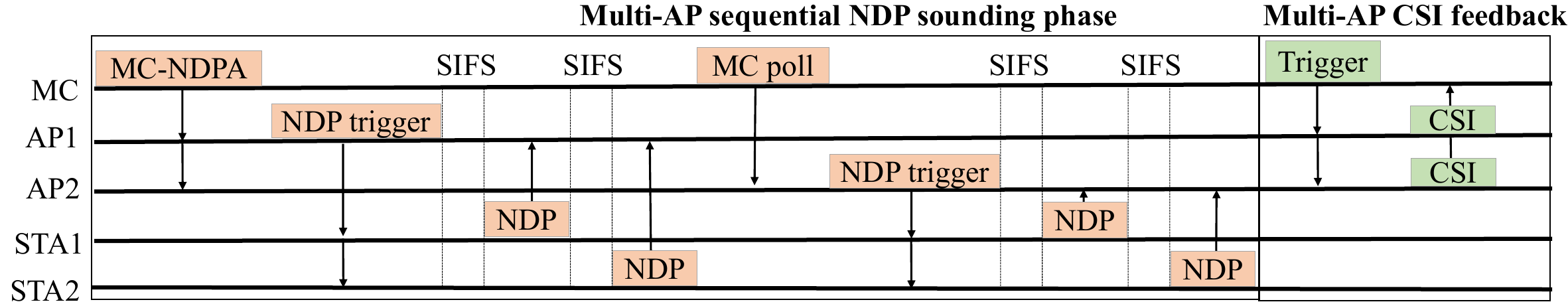}
	\caption{Multi-AP MC-based implicit sequential channel sounding.}
	\label{MC-implicit}
	\vspace{-0.5cm}
\end{figure*}
   \begin{figure*}
	\centering
	\includegraphics[width=7in]{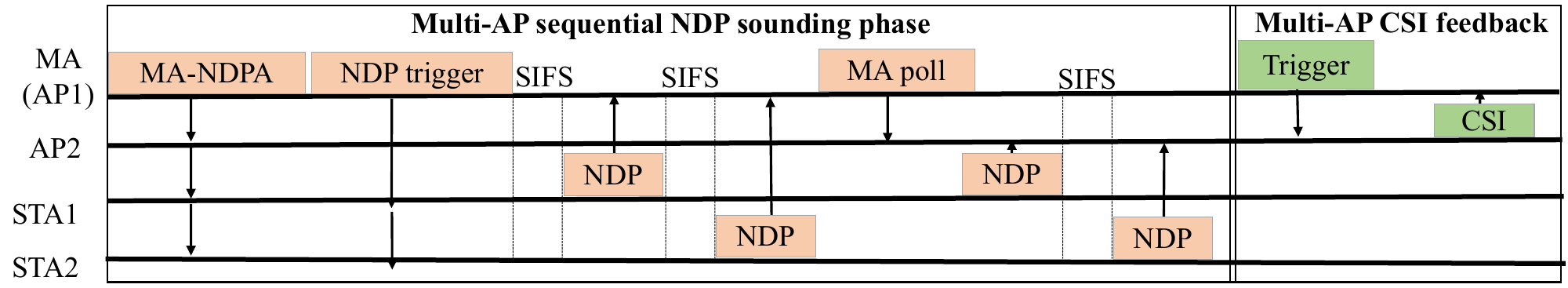}
	\caption{Multi-AP MA-based implicit sequential channel sounding.}
	\label{MAP-implicit}
	\vspace{-0.5cm}
\end{figure*}
It is important to note that EHF bands, like 802.11ay, do not have a specific MAP-Co channel sounding method outlined in their protocols. Instead, 802.11ay's MAC protocol divides the beacon interval into two parts: the beam header interval and the data transmission interval. The protocol uses beam training and sector sweeping to optimize beamforming and understand the channel quality of STAs of an AP. CSI feedback is reported during beamforming training and beam tracking through beacon transmission and sector-level sweeping (SLS). The beacon interval structure and beam training frame MAC structure for a single AP 802.11ay protocol are explained~\cite{aysurvey,ieee4}. Other WLAN protocols should consider enabling MAP-Co explicit channel sounding in their MAC protocols.

 \subsection{Implicit Channel Sounding}
 To provide feedback in a network with a large number of STAs and antennas at AP, the explicit channel sounding process can increase network overhead and introduce delays before data transmission. To address these issues, researchers have proposed the implicit channel sounding method. This method relies on channel reciprocity and uses CSI at the transmitter to calculate CSI at the receiver. It estimates CSI from a null signal transmitted by STAs and estimates downlink (DL) CSI based on received uplink (UL) channel sounding. The implicit sounding assumes that the impulse response of the uplink (UL) and downlink (DL) channels are identical within the same coherence interval. Several studies, such as \cite{sch,gast,doo,roga,shepard}, have examined this method and its advantages over explicit sounding.
In reality, the UL and DL baseband channels are not the same due to non-reciprocal impairments that differ in the baseband-to-RF and RF-to-baseband chains. To adjust for this difference, a calibration approach is used. Implicit channel sounding is a method that provides lower overhead and latency as it doesn't require beamforming feedback information or quantization. In this method, an AP sends NDP trigger packets to STAs and each STA responds with NDP packets to the AP with a gap of SIFS frames between each response. The AP estimates the CSI based on the received NDP packets from STAs. The UL channel sounding can be used as DL channel sounding for UP since both have the same impulse response in the same interval. However, a modified single AP implicit channel sounding is used for the MAP-Co environment.

When dealing with MAP-Co over 802.11be, the implicit channel sounding approach needs to be adjusted based on MC and MA-based architecture. The IEEE standard, as described in \cite{ryu}, discusses this approach for MAP-Co. To initiate implicit channel sounding for MAP-Co, MC or MA triggers slave APs and STAs to transmit NDPA/NDP frames. Additionally, there are two types of multi-AP implicit channel sounding approaches: MC-based and MA-based. These two types can also be further divided based on the NDP transmission pattern by STAs, which are sequential and joint. To illustrate, Fig.~\ref{MC-implicit} shows the MC-based implicit sequential channel sounding approach, while Fig.~\ref{MAP-implicit} presents the MA-based sequential implicit channel sounding.

In Figure ~\ref{MC-implicit}, the MC broadcasts NDPA trigger frames to all slave APs, which includes the ID of all slave APs and user information for its STAs and other STAs in OBSS. The user information consists of resource units on which NDP frames would be transmitted, multiplexing type for multi-user NDP, number of antennas to transmit NDP frames and more. After receiving the trigger frame, slave APs transmit NDP trigger frames to all STAs. In sequential implicit channel sounding, STAs transmit NDP frames to their APs in the order mentioned in the NDP frames, after a SIFS interval since receiving the NDP trigger frames. There must be a gap of SIFS between NDP frames transmission by each STA. In joint implicit channel sounding, STAs transmit NDP frames together in different resource units. After receiving the NDP frames, slave APs perform calibration to estimate the channel states. Thereafter, the MC sends MA trigger frames to slave APs to report the measured channel states, and slave APs report their estimated channel reports together.

In the MA-based implicit channel sounding method, the MA sends NDP trigger frames to its STAs and slave APs as shown in Fig.~\ref{MAP-implicit}. The slave APs then send the same frames to their own STAs. The STAs respond to the trigger frames with NDP frames either sequentially or together. In sequential MA-based implicit channel sounding, the STAs of the MA respond with NDP frames one after the other with a SIFS interval. The STAs of the slave APs then transmit NDP frames to their own APs sequentially and with a SIFS interval. In joint MA-based implicit channel sounding, all STAs of the MA and slave APs respond with NDP frames at the same time. This article provides a taxonomy of implicit channel sounding based on MAP-Co architecture and NDP frame transmission patterns. In the next subsection, we will discuss the summary and lessons learned from the discussion of MAP-Co implicit channel sounding.

  \subsubsection{Summary and Lesson Learned}
Table~\ref{sounding-comparision-table} summarizes the various types of implicit channel sounding used for MAP-Co, along with their advantages and drawbacks. By studying these methods, we have learned valuable lessons. Among the different methods, the MC-based sequential implicit channel sounding method requires less calibration processing and does not congest the UL network. This approach has sufficient computation resources, thanks to high-end servers, which reduces the processing delay during calibration estimation. However, STAs may have to wait longer to transmit the NDP frames to their slave APs with this method, and it does not properly utilize OFDMA-based network resources for transmission by multiple STAs simultaneously. The MC-based joint implicit channel sounding method, on the other hand, efficiently utilizes the UL network resources for transmission by multiple STAs during NDP frames, and it can perform multi-channel sounding for an STA. In this method, the calibration estimation processing for multiple NDPs at the same time is not an overhead due to high-end servers. However, this method can increase the network overhead if there are numerous STAs per AP. Furthermore, machine learning approaches for channel estimation can also be used instead of implicit channel sounding.

When using MA-based sequential implicit channel sounding, the amount of communication required is lower compared to sequential MC-based, as MA can send trigger frames to multiple STAs and slave APs simultaneously, reducing the number of exchanges needed. This method also experiences lower NDP transmission delay and has no UL network congestion. However, MA can experience high calibration calculation overhead and longer processing delays due to limited processing power, so it may not be suitable for a large number of devices per AP. Additionally, like other sequential methods, it may not efficiently utilize OFDMA-based network resources. The joint MA-based implicit channel sounding can use network resources efficiently with lower waiting time for transmission of NDP frames, but it may significantly increase the overhead at MA due to limited processing resources, leading to inaccurate channel estimation and high latency. This method can also congest UL channels during reporting of NDP frames, so it may not be suitable for MAP-co in high-device scenarios. Therefore, there is a need for an approach that can optimize communication and computation overhead of joint MA-based implicit channels sounding together.

The quality of implicit sounding in IEEE 802.11be depends on the reciprocity nature of both DL and UL channels. However, this may not be suitable for the 802.11ay spectrum due to non-reciprocal interference from other APs. 802.11ay APs have a small coverage area, which can cause signal interference with neighboring STAs and APs. Additionally, the implicit approach is not effective on OFDM-based channels because of their non-reciprocal nature. To overcome these issues, researchers have studied implicit channel sounding, which is commonly used in designing hybrid analog-digital beamforming~\cite{cs1, xiwen}. Also, various calibration methods have been proposed for mmWave WLAN, but they are mainly for single-AP 802.11ay WLAN \cite{cal1,cal2}. In the future, these issues should be studied for MAP-Co in 802.11ay WLAN. It is important to note that implicit channel sounding in 802.11ay WLAN may experience similar overhead as explicit due to the large number of slave APs in an area compared to 802.11be WLAN. Therefore, an overhead analysis needs to be conducted for implicit channel sounding in 802.11ay WLAN.


 
\section{MAP-Co Based Scheduling Approaches}
In this section, we will take a look at the latest MAP-Co features that allow for efficient resource allocation among multiple APs and simultaneous transmission by multiple APs. To start, we'll give an overview of C-OFDMA and the current research on this technology, including a summary of key findings and lessons learned. Next, we'll delve into CSR operations, exploring the various issues that have been studied in this area and summarizing our findings. We'll also explain the process of achieving CBF, discuss the current state-of-the-art in CBF, and share what we've learned through our research. Lastly, we'll provide an explanation of JTX operation, highlight some of the existing literature on this topic, and summarize our key takeaways.
\label{maptransmission}
  \subsection{Coordinated OFDMA}
\begin{figure}[t]
	\centering
	\includegraphics[width=3.5in]{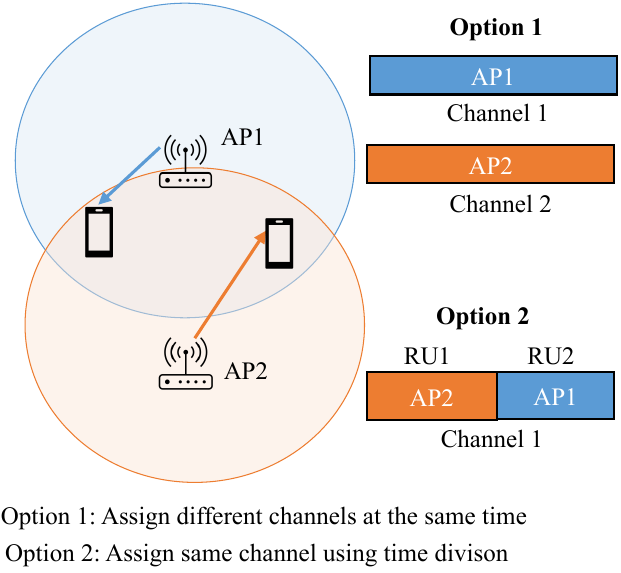}
	\caption{Basic concept of C-OFDMA.}
	\label{cofdmafig}
	\vspace{-0.5cm}
\end{figure}

OFDMA is a significant feature of the 802.11ax/be WLAN, which allows multiple STAs to transmit/receive data simultaneously by sharing bandwidth, thereby improving throughput and reducing latency. In OFDMA, the bands are represented in terms of sub-carriers, which are grouped to form resource units (RUs). 
In a single AP environment, an AP cannot access the same frequency as another neighboring AP in the same TXOP, limiting the system's ability to achieve higher throughput. To address this limitation, C-OFDMA has been proposed to enable sharing of OFDMA resources among APs and STAs at the same TXOP while avoiding RU conflicts and interference. For instance, two STAs close to their BSSs can access the same frequency or RUs if they are coordinated, as they cannot interfere with each other. Moreover, edge STAs cannot transmit on the same RUs and frequency as they can interfere with each other. As shown in Fig.\ref{cofdmafig}, C-OFDMA allows for the assignment of different RUs in the same frequency or different frequencies for transmission to edge STAs at the same TXOP.
However, in the upcoming 802.11be, a group has been formed to discuss C-OFDMA operation in 802.11be-enabled APs~\cite{cofdma6,cofdma}. Therefore, in this article, we present the discussed C-OFDMA operations for 802.11be protocols.

  \begin{figure}[t]
	\centering
	\includegraphics[width=3.5in]{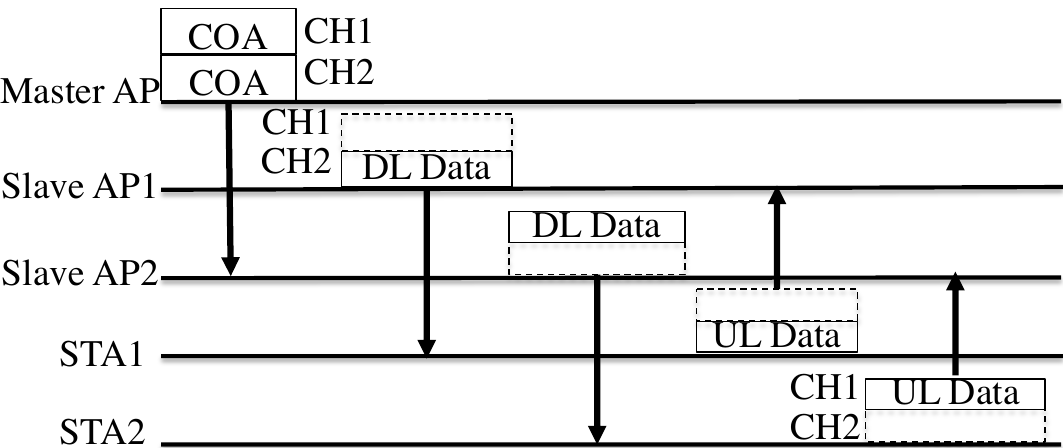}
	\caption{C-OFDMA operations.}
	\label{ofdma-operation}
	\vspace{-0.5cm}
\end{figure}
\subsubsection{C-OFDMA Operation}
The MC/MA is responsible for directing the slave APs on how to access frequency resources in C-OFDMA, which consists of three steps: C-OFDMA setup, transmission scheduling, and transmission. During the C-OFDMA setup, the MA/MC shares available resources with all slave APs and receives requests from them indicating their desire to participate by specifying the required resources (frequency or time). These requests can be included in the physical layer protocol data unit for single-user or high-efficiency trigger-based PPDU for UL MU transmission if supported by the slave APs. The MA/MC then schedules the C-OFDMA transmission by allocating bandwidth, length, and other transmission parameters (TXVECTOR) for both DL and UL transmission. The TXVECTOR defines the maximum length of MPDU, data rate, and other relevant details. The MA/MC shares the allocation information with the slave APs. As shown in Fig.\ref{ofdma-operation}, it is possible to transmit two COA frames over two separate channels. One COA frame is sent to AP1 via channel CH1, while the other is transmitted to AP2 via channel CH2. STA1 is associated with AP1, and STA2 is associated with AP2. In this setup, the MA/MC does not allocate RUs to STAs. Instead, the available RUs of the allocated channels are assigned by the slave APs to their respective associated STAs. For example, AP1 can assign RUs of CH1 to STA1, and AP2 can allocate RUs of CH2 to STA2. Once the RUs are assigned, STA1 transmits UL data on the assigned RUs of CH1, and STA2 transmits UL data on the RUs of CH2. In case of DL transmission, the slave APs transmit DL data to their associated APs on the channels where they receive COA frames, as depicted on the right side of Fig.\ref{ofdma-operation}. Hence, C-OFDMA can prevent collisions between neighboring APs transmitting over the same RUs by coordinating their transmissions. However, if slave APs choose the same channels before C-OFDMA, they may receive COAs in the same primary channels, which can cause collisions. To avoid this, the coordinator can instruct the slave AP to switch to another primary channel within the allocated bandwidth or assign different RUs to slave APs within the same channel. With C-OFDMA, different slave APs can have different primary channels within the allocated bandwidth. However, using the same primary channel can lead to ICI issues. Therefore, coordinated RUs and multiple channel allocation among different APs are important research areas for C-OFDMA to prevent collisions and interference. In the next section, we will provide a comprehensive overview of the research areas of C-OFDMA.
\subsubsection{State-of-the-art of C-OFDMA}
In the area of C-OFDMA, researchers have been working on ways to manage WLAN resources in order to improve the coordinated TXOPs of multiple devices and maximize throughput. This section will present the latest WLAN OFDMA resource management schemes by multiple coordinated APs. As mentioned earlier, there are two types of resource allocation among different APs: using the same channel but different RUs at neighboring APs, or using different channels for each neighboring AP. The first approach may cause co-channel interference issues, while the second method may introduce inter-channel interference (ICI) issues. Therefore, C-OFDMA-based resource allocation strategies can be categorized into two groups: inter-channel and intra-channel assignment. 

\begin{table*}[!t]
	\renewcommand{\arraystretch}{1.3}
	\caption{State-of-the-art of C-OFDMA.}
 \label{cofdma}
\begin{tabular}{|m{2.7cm}|m{1.4cm}|m{4.4cm}|m{3.2cm}|m{4.4cm}|}
\hline
C-OFDMA Taxonomy                & Related Work                      & Main features  &objective                                               & Drawbacks  \\ \hline
\multirow{5}{*}{Inter-Channel Allocation} & \cite{cofdma1}             & Multi-cell C-OFDMA resource management approach by selection of optimal sets of sub-channels sequence for each cell and scheduling of each set.                                                                                        & Reduce ICI and improve spectral efficiency                                                      & Use brute force to determine different combinations of sub-channels; higher complexity in case of large number of coordinated APs.                                                                               \\ \cline{2-5}
                                          & \cite{cofdma2,cofdma3} & Sub-channel assigning scheme for DL C-OFDMA networks proposed. Based on graphs, uses MAX-k-cut graph coloring algorithm while considering location and channel information.                                                              & ICI and improve spectral efficiency                                                                                       & Overlooked addressing real-time channel quality data collection and scalability concerns for mobile devices.                      \\ \cline{2-5}
                                          & \cite{cofdma4}            & Proposed a method called the centralized iterative water-filling algorithm that efficiently allocates sub-channels for the DL C-OFDMA network.                                                                                                              & Coordinated power control for maximizing sum rate.                                                                       & Perfect synchronization between APs and fixed channel gain information for sub-channel selection is not possible in emerging mmWave WLAN due to fast and unpredictable channel changes.\\ \cline{2-5}
                                          & \cite{cofdma5}           & Proposed a method for allocating resources in 3D antenna array systems that use OFDMA technology. The approach involves splitting the downltilts between cell-center and cell-edge users using vertical dynamic beamforming.        & Maximize throughput and reduce ICI                                                                                        & Complete avoidance of Inter-Cell Interference (ICI) is not achievable if the power allocation of center-cell and cell-edge STAs is not shared with other APs.                                                                                                      \\ \cline{2-5}
                                          & \cite{cofdma6}          & 
Suggested a framework for the MAC layer that facilitates MAP and multi-band operations, enabling the allocation of distinct bands to each AP through coordination.                                                                                                          & Increase throughput and minimize ICI                                                                                                           & The full potential of C-OFDMA cannot be reached because the entire band is reserved for an AP.    \\ \cline{2-5}
& \cite{laca}         & 
Proposed coordinated resource units allocation for time-sensitive applications.                                                                                                        & Maximizing the probability of successful data delivery.                                                                                                           & Analyzed only for small networks; performance may be limited for larger ones.
\\ \hline
\multirow{6}{*}{Intra-Channel Allocation} & \cite{cofdma7}        & In DL MIMO OFDMA, multiple BSSs work together using coordinated linear precoding to minimize co-channel interference.                                                                                                             & Weighted sum rate maximization, and maximizing the weighted sum of the minimal user rates of the coordinated BSSs & The estimation of precoding matrix for each AP-STA pair is computationally expensive, requiring significant computational resources.                                                                                                                          \\ \cline{2-5}
                                          & \cite{cofdma8}          & The suggested precoding method, termed dirty paper coding, requires the encoder for the current user to be aware of the encoding schemes, associated CSI, and propagation delays used by previous users. & Reduce co-channel interference                                                                                           & Higher signal overhead and preparation delay before precoding.                                                                                                                                                    \\ \cline{2-5}
                                          & \cite{cofdma9}             &The interference reduction was achieved by optimizing both the precoder and postcoder through the design of a coordinated precoder and postcoder.                                                                                                                          & Reduce co-channel interference and minimized errors                                                                       & The iterative approach becomes a potentially NP-hard problem, particularly when dealing with a larger number of coordinated APs.                                                                                                                 \\ \cline{2-5}
                                          & \cite{cofdma10}         & Proposed energy-aware downlink resource allocation method, optimizing user scheduling and power allocation across coordinated base stations while accounting for constraints on transmit power per allocated RUs in the same channel.     & Minimize energy and maximize spectral efficiency                                                                          & Achieving perfect synchronization and accurate CSI is not a realistic scenario in the evolving WLAN environment, posing a significant challenge.                                                                                \\ \cline{2-5}
                                          & \cite{cofdma11}            & The proposed approach involves the selection of users for co-channel access in each tone and determining the power allocation across the different tones.                                                                                                                               & Maximize the weighted system sum rate                                                                                                          & The combinatorial optimization approach may encounter elevated time complexity, especially when dealing with a large number of STAs and APs.                                                                                          \\ \cline{2-5}
                                          & \cite{cofdma12,cofdma13}                                  & Proposed mathematical models to analyze the trade-off between spectral and energy efficiency in UL C-OFDMA.                                                                                                                                                       & Improve energy and spectral efficiency.                                                                                                        & The proposed method did not consider co-channel interference.                                                                                                                                                                        \\ \hline
\end{tabular}
\end{table*}
i) \textit{C-OFDMA based Inter-channel  Allocation:} C-OFDMA resource allocation often utilizes inter-channel resource allocation to mitigate co-channel interference between different basic service sets (BSSs) or cells. This method assigns different primary channels to different BSSs to avoid interference, but it can introduce ICI. To tackle this challenge, Tim et al. \cite{cofdma1} proposed a multi-cell C-OFDMA resource management scheme. They allocated specific sub-channel sequences to each type of cell, resulting in a significant improvement in spectral efficiency, particularly for sector-wise cells. By allocating a subset of sub-channels based on user requests, collisions between neighboring cells can be minimized. In their conclusion, the authors found that sector-wise cell allocation significantly enhances spectral efficiency in both sector-wise and omnidirectional cells. They also noted that this approach can be applied to emerging WLAN technologies, including sector-wise beamforming in mmWave WLANs. 

In another study by different authors, a graph-based sub-channel assignment scheme was proposed for downlink (DL) OFDMA networks \cite{cofdma2, cofdma3}. The scheme consisted of two phases: location-aware interference management and channel-aware sub-channel assignment. The authors utilized graph coloring techniques, where STAs were represented as nodes and ICI as edges. Taking into account the location and movement of STAs, the MAX-k-cut algorithm was employed to identify disjoint clusters with reduced interference. Sub-channels were then assigned to these clusters based on the instantaneous channel quality. However, the study lacked detailed information on obtaining instantaneous channel quality and scalability challenges could arise with a larger number of mobile devices. Additionally, the model did not consider different mobility patterns.
The article \cite{cofdma4} presents a method for efficiently assigning sub-channels to DL C-OFDMA networks using a centralized iterative water-filling algorithm. The authors also suggest coordinated power control to maximize the overall rate. This involves adjusting the power levels for each base station and calculating the sum rate at each iteration until it is optimized. However, it is important to note that this approach assumes that there is perfect synchronization between base stations and that channel gain information is available for each user and base station. This may not be practical for emerging WLANs where the network environment and cells cause a constantly changing channel quality, requiring continuous synchronization and channel gain information.
Zhang et al. \cite{cofdma5} presented a resource allocation plan for 3D antenna array systems that optimizes physical layer parameters. They utilized vertical dynamic beamforming to allocate downtilts for cell-center and cell-edge users. Their aim was to enhance throughput for both types of users by coordinating transmission power and downtilt adjustments based on assigned sub-channels. To prevent interference during transmission by cell-center users, their approach assigns RUs and adjusts physical layer parameters for cell-edge users. However, there is limited research on inter-cell resource allocation over WLAN using C-OFDMA.

 In this vein, the authors introduced a MAC layer framework for C-OFDMA over multi-band WLAN \cite{cofdma6}. To avoid interference, they recommended assigning different frequency bands to each access point (AP) through coordination. However, their approach reserves entire bands for all stations (STAs) and their APs, even when there is no overlap or interference. As a result, this approach may not be efficient for real-time or industrial applications. To allocate resources more efficiently to different sub-channels, it is crucial to consider the specific requirements of each application. A recent article proposed a method called C-OFDMA that utilizes a frame exchange scheme based on implicit wireless channel sounding before transmission \cite{laca}. The scheduler assigns resource units to users to increase the likelihood of successful data delivery for industrial Wireless Time-Sensitive Networking scenarios. Additionally, a virtual sounding mechanism is suggested to evaluate channel quality without causing network overhead.
 Another approach is C-OFDMA-based intra-channel allocation, which is presented below.

ii) \textit{C-OFDMA based Intra-channel Allocation:} In Intra-channel assignment, neighboring BSSs allocate the same channel to different users. The main challenge is to prevent co-channel interference while maximizing network throughput and other performance metrics. There are three methods to efficiently assign RUs within the same sub-channels in case of OBSSs while avoiding interference. The first two methods optimize physical layer parameters, including precoding and transmission power control (TPC). The third approach involves improving the WLAN's MAC layer by using a grouping and graph-based method.

The work by Wang et al. \cite{cofdma7} proposes the use of coordinated linear precoding in DL MIMO OFDMA networks to reduce co-channel interference among multiple BSSs. The authors focus on two precoding design problems: weighted sum rate maximization and maximizing the weighted sum of the minimal user rates (MWSMR) of the coordinated BSSs while considering power constraints per cell. They propose mathematical models to determine the precoding matrix for a certain number of users of multiple BS accessing a sub-channel, such that interference can be avoided. Linear precoding is used because signals of other users can be treated as noise at the intended user. In a similar vein, another approach to avoid co-channel interference in a coordinated way was proposed in \cite{cofdma8}. This approach is based on precoding and is called dirty paper coding. With this method, the encoder for the current user requires knowledge of the encoding of previous users and associated CSI and propagation delays to cancel out any interference caused by those users. Kasaeyan et al. \cite{cofdma9} proposed a coordinated precoder and postcoder design approach to reduce interference in DL C-OFDMA networks. The idea is to fix one coder and optimize the other, with the process being repeated until the error is minimized. Their formulation is specifically for DL and their proposed approach improves throughput significantly compared to single-cell optimization. Additionally, other works have also utilized TPC for mitigating co-channel interference in C-OFDMA. The authors of \cite{cofdma10} proposed an energy-aware approach for downlink resource allocation in C-OFDMA, which optimizes user scheduling and power allocation across a group of coordinated base stations. The approach aims to balance the trade-off between energy efficiency and spectral efficiency while ensuring a constraint on the transmit power per allocated RUs in the same channel. The optimal number of users from multiple cells assigned over sub-carriers and the transmit power level were determined to achieve maximum efficiency. However, the approach assumed perfect synchronization and accurate CSI, which may not be feasible in practical WLAN scenarios as previously discussed. An additional study has also addressed the challenge of co-channel interference in C-OFDMA networks by selecting users for co-channel access in each tone and determining the power allocation across tones to optimize the weighted system sum rate for DL transmission \cite{cofdma11}. While most of the existing literature has focused on the DL C-OFDMA, there are some studies on the UL C-OFDMA. For example, in \cite{cofdma12}, the authors developed mathematical models to investigate the trade-off between spectral and energy efficiency in UL C-OFDMA. They demonstrated that their proposed energy models for C-OFDMA can result in significant energy savings. Therefore, coordinated TPC over each RU or sub-carriers can be an effective way to avoid interference and enhance both energy and spectral efficiency. The authors of \cite{cofdma13} proposed an energy-efficient sub-channel allocation scheme with target wake-time scheduling for multiple BSSs using a MAC-based C-OFDMA intra-channel allocation approach. Their method involved dividing STAs under coordinated APs into groups and creating an undirected graph of OBSS where vertices represented STAs and edges represented overlapping STAs. They assigned different colors to vertices that were not adjacent to each other and then used the Welch-Powell method to find the minimum number of colors required to maximize the number of parallel transmissions while avoiding interference. They proposed a scheduling approach to determine the transmission and sleep times for each group, with each group consisting of vertices of the same color to enable concurrent transmission without interference. This is considered one of the best MAC-layer-based approaches for implementing C-OFDMA-based intra-channel allocation.

\subsubsection{Summary and Lesson Learned}
C-OFDMA resource allocation can be divided into two types: inter-channel and intra-channel. Inter-channel allocation coordinates multiple channels among different APs to minimize inter-cell interference, while intra-channel allocation assigns different RUs of the same channel to STAs of different APs through coordination. Table~\ref{cofdma} provides an overview of the state-of-the-art for both types of C-OFDMA. Inter-channel allocation typically uses combinatorial optimization to determine the best sub-channel sets for each BSS, but this method can be time-consuming. Alternatively, graph coloring algorithms can be used to optimize sub-channel sets. Most approaches aim to reduce inter-cell interference, but they require perfect synchronization and knowledge of CSI, which may not be practical in mmWave WLAN due to environmental dependencies.
To allocate resources within a channel, there are different approaches such as coordinated precoding, coordinated TPC, and coordinated MAC-layer based methods. However, coordinated precoding has a longer preparation delay and higher signal overhead, while coordinated TPC requires perfect synchronization, which can be difficult for a large number of STAs. Additionally, some approaches involve determining combinations of STAs and RUs to avoid interference, but this can be time-consuming and require complex algorithms to find the best sets. Therefore, there is a need for a more efficient and proactive C-OFDMA resource allocation approach.
\begin{figure}
	\centering
	\includegraphics[width=3.5in]{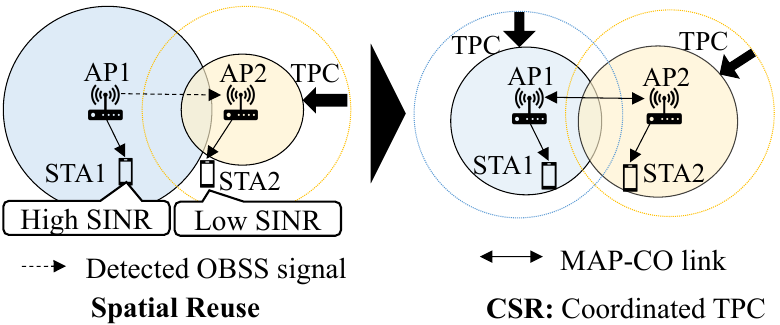}
	\caption{The difference between SR and CSR.}
	\label{csr-example}
	\vspace{-0.5cm}
\end{figure}

\subsection{Coordinated Spatial Reuse}
Spatial reuse (SR) is a method introduced by the 802.11ax standard to increase parallel transmission and spectral efficiency in dense WLAN environments through transmission power control and modulation and coding schemes (MCS). In this method, the Overlapping Basic Service Set (OBSS) Packet Detect (PD) technique adjusts the clear channel assignment/carrier sense (CCA/CS) threshold for the detected OBSS to the OBSS packet detects (OBSS-PD) threshold~\cite{ieee8}. One AP is chosen to enable OBSS-PD threshold while other APs can use the CCA/CS threshold. This ensures that transmissions at different thresholds do not overlap, resulting in improved TXOP and spectral efficiency. However, some STAs may experience low SINR when an AP decreases its transmission power, and coordination is necessary for dense AP environments to avoid interference. To address this issue, the 802.11be standard introduced CSR to perform SR in a coordinated manner. CSR enables APs to perform coordinated TPC to maintain adequate SINR at each STA, as shown in Fig.~\ref{csr-example}. Different aspects of existing CSR operations are described, including architecture, preparation stage, data transmission procedure, and control information exchange between APs.

\subsubsection{CSR operations}
The CSR architecture is based on the MAP-Co architecture. If using an MC-based MAP-Co, only slave APs are involved in CSR. If using an MA-based MAP-Co, there are two options for CSR. Option 1 is to have a fixed AP as the coordinator and all other APs as slaves. Option 2 is a dynamic approach where the AP that wins TXOP first becomes the coordinator and other APs willing to be coordinated become slaves.

In CSR, the transmission power of each AP is determined by the coordinator (MC/MA) based on information obtained from channel measurement reports and beacon frames. The coordinator estimates the receiving signal strength index (RSSI) of STAs at each channel and the mutually acceptable receiver interference level between neighboring APs using the measurement reports. The coordinator then uses this information to calculate the transmission power of each AP for its OBSS-PD threshold. The transmission power for the OBSS-PD threshold of a slave AP can be estimated using an equation that takes into account the difference between the desired OBSS-PD threshold and the CCA/CS threshold, as well as the difference between the RSSI and noise floor, as mentioned in equation~\ref{csr-eq}~\cite{csrproc}. This calculation ensures that the SINR of STAs is maintained at an adequate level, even when the APs are transmitting at lower power levels.

\begin{align}
\label{csr-eq}
    TX\_PW\_AP1 &\leq TX\_PW\_NAP \nonumber \\
    &\quad - RSSI\_AP1 + ARIL\_N
\end{align}

Let's focus on AP1 as the slave AP that requires adjustment of the CCA/CS threshold and estimation of maximum TP. In the equation above, $TX\_PW\_AP1$ represents the average transmission power of AP1, which needs to be OBSS-PD. $TX\_PW\_NAP$ represents the transmission power of neighboring slave APs (NAP), such as AP2. $RSSI\_N$ represents the RSSI values for a signal from AP1 to the STA of the neighboring slave AP, and $ARIL\_N$ represents the acceptable receiver interference level (ARIL) at the STAs of NAP. After estimating the new TP of an AP, the ARIL values of STAs of AP1 ($ARIL-AP1$) should be updated. AP1 can only select an STA within its new range when the updated $ARIL$ of STAs is greater than the RSSI from NAP to the STAs, which shows interference control through CSR. The updated $ARIL-AP1$ is equal to ($TX\_PW\_AP1-pathloss-min SNR-safety margin$). To provide an example of the estimation of TP in CSR, let's consider two slave APs (AP1 and AP2) and an MC, as shown in Fig.~\ref{csr-example}. STA1 is associated with AP1, and STA2 is associated with AP2. The dotted circle in Fig.~\ref{csr-example} presents the original logical coverage of AP1 and AP2, as well as the distance between STAs and each slave AP. When AP2 transmits a packet to STA2 in the same TXOP, STA1 can experience interference and may find that the channel is not idle. However, STA1 is far from AP2, so AP2's transmission to STA2 cannot impact STA1. To perform CSR, the maximum TP of AP2 should be decreased to reduce the coverage during that TXOP. The updated $ARIL$ of STA1 and STA2 should be estimated, and the updated $ARIL$ of STA1 will be greater than the $RSSI$ from AP2 due to the out-of-coverage area of AP2 and lower TP. The 802.11be proposed two options of CSR based on the timing of decision-making. In option 1, the TPC is performed periodically by each slave AP, and the coordinator decides the TPC and other settings periodically. The coordinator informs slave APs through periodic transmission of CSR trigger frames, which carry control information such as the identity of slave APs, PPDU length of CSR data transmission, length of the basic time frame, maximum TP of each slave AP, the TX power of the coordinator for transmission by slave APs, etc. This approach does not increase signal overhead during CSR process. The other option is to control the TP in every transmission. The coordinator exchanges CSR trigger frames before each transmission, and slave APs adjust their CCA/CS threshold accordingly by TPC. Option 2 increases system overhead with more CSR trigger frames assignment, but it is dynamic and improves the system throughput at each transmission. Option 2 is highly suitable for mobile STAs. In the next section, we will summarize the available literature addressing the issue of CSR.
\begin{table*}[!t]
	\renewcommand{\arraystretch}{1.3}
	\caption{State-of-the-art of different methods of CSR.}
	\label{table-csr}
\begin{tabular}{|m{2.1cm}|m{1.8cm}|m{5.9cm}|m{1.7cm}|m{4.5cm}|}
\hline
Taxonomy                 &Related Work& Main features          & Approach type            &Drawbacks  \\ \hline
\multirow{4}{*}{OBSS-PD CSR} & \cite{CSR1} & The proposal aims to efficiently utilize available links by determining coordinated links and MCS for ongoing transmission and SR transmission with OPCSR. &Transmission-wise TPC of slave APs  & Many data streams can cause delays in synchronization and communication due to waiting for available links.  \\ \cline{2-5} 
                        & \cite{CSR2} & Analyzing OPCSR performance for two links with varying TPC and MCS settings on two slave APs in CSR. &Transmission-wise TPC of slave APs  & This proposal pertains to a small scenario that may not be very realistic for the emerging WLAN technology  \\ \cline{2-5}  
                          & \cite{CSR3} & The authors developed a mathematical model that suggests interference-free channels and assigns them to avoid any mutual interference. &Periodic  & Statistical model; not fit for realistic dynamic emerging WLAN.  \\ \cline{2-5} 
                          & \cite{CSR4} & Proposed integrated CSR and C-OFDMA based interference management. &Transmission-wise TPC of slave APs  &No MAC framwork to integrate different MAP-Co transmission approach.   \\ \hline
 \multirow{4}{*}{Parameterized CSR} & \cite{CSR6} & The authors have presented a mathematical model that schedules joint and equitable transmission, along with an algorithm for adapting the power spectrum. &Transmission-wise TPC of slave APs  & Designed for cellular networks.  \\ \cline{2-5} 
                        & \cite{CSR7} & The authors analyzed the performance of PSCR and coordinated scheduling to improve spatial reuse in cellular networks.   &Transmission-Wise CSR & Studied for cellular networks; Not applicable for unlicensed spectrum \\ \cline{2-5}  
                          &   \cite{pscr-wlan} &The authors have proposed a Q-learning based approach to share information about the schedule of access points that should be involved in CSR over WLAN. &Periodic CSR  &   No approach to mitigate inter-cell interference\\ \cline{2-5} 
                           &   \cite{chem} & Proposed approach ensures efficient PCSR utilization for real-time applications STAs while maintaining performance for non-real time application STAs.  & Periodic CSR  &    Analyzed for a network with up to 6 STAs; performance may degrade with more.
                            \\ \hline
\multirow{3}{*}{MAC-based CSR}   & \cite{CSR8} & A novel MAC framework based on grouping has been suggested for implementing CSR by sharing TXOP in 802.11be.&Transmission-wise CSR  & The proposed approach only takes into account the static STAs and WLAN environment, which may not be effective in a dynamic WLAN environment where there are new AP deployments and moving STAs. \\ \cline{2-5} 

& \cite{CSR9} & To prevent interference and increase network speed, mathematical models were created to analyze the effects of combining slave APs and STAs during a TXOP. The chosen combination should result in minimal interference and maximum throughput.  &Periodic CSR& To find combinations, a brute force algorithm is used. However, there is a lack of analysis on the interval required to find these combinations. This method results in a higher overhead for a TXOP. \\ \cline{2-5} 

& \cite{david} & The study focuses on creating multi-AP groups for simultaneous transmission and evaluates per-AP and per-group scheduling algorithms based on individual and collective buffer states, respectively.  &Periodic CSR& Studying 9 APs with a maximum of 3 per group underscores the necessity for future analysis in larger networks with a higher density of APs.

\\ \hline
\end{tabular}
\end{table*}
\subsubsection{State-of-the-art of CSR}
In this section, we will provide a breakdown of the different approaches to CSR and discuss the current issues in each category. There are three main categories of CSR approaches: OBSS PD-based CSR (OPCSR), Parameterized CSR (PCSR), and MAC-layer-based CSR (MCSR).

\textit{OBSS PD-based CSR:} The OPCSR approach regulates the transmission power of an OBSS to ensure that the RSS of any ongoing transmissions from a neighboring cell is lower than the OBSS-PD threshold of the OBSS~\cite{CSR}. This allows for controlled OBSS to begin transmitting without interference from neighboring cells. Lee et al.~\cite{CSR} researched situations of mutual interference, even when RSS values are below the OBSS threshold. They found that failed transmission is possible if the ongoing frame receiver is close to SR transmission on the same link, even after OPCSR. Therefore, they suggested determining coordinated links for ongoing transmission and SR transmission with OPCSR, so that available links can be used efficiently. They also proposed determining coordinated MCS values for SR and ongoing links. This distributed approach has a lower overhead but may have synchronization and higher communication delays in the case of a larger number of data streams, leading to longer waiting times for available links. Thus, this approach is suitable for a few CSR-based DL data streams but not ideal for CSR-based UL transmission. It is worth noting that the proposed approach did not investigate the limitations of link-aware OPCSR for more than two slave APs.
Similarly, in \cite{CSR2}, simulations were conducted to assess the performance of OPCSR in various scenarios involving two slave APs. The authors analyzed path loss between APs and STAs and altered TPC and MCS to determine throughput, packet error rate, and optimal MCS values. This analysis can assist researchers and industries in determining the best MCS values for CSR in different circumstances. However, the study was conducted on a small scale, which may not be entirely realistic, and scalability could be a concern when implemented in a real network. 
Also, a new method has been suggested to enhance the quality of links when using OPCSR with uncorrelated antennas, as discussed in \cite{CSR3}. The authors proposed a mathematical model that identifies channels that are free from interference and assigns them accordingly to prevent mutual interference. However, this statistical model is only suitable for DL loads and cannot be used with UL CSR or real WLAN. Most of the existing research on OPCSR has focused on DL transmission. Another way to manage interference with OPCSR is to integrate it with C-OFDMA, as explained in \cite{CSR4}. In this integrated system, coordinated APs share assigned RUs with their cell-edge STAs, and neighboring APs determine transmission power for transmission at RUs accordingly.

\textit{Parameterized CSR:} The PCSR approach involves the coordinator informing multiple STAs of different slave APs about spatial reuse TXOP through a trigger frame. This frame contains information about transmission scheduling, allowable interference levels, and transmit power of other slave APs to estimate interference. PCSR has several benefits, including reducing contention time during CSR, allowing for more concurrent TX through high resource utilization compared to OPCSR, and reducing latency for time-sensitive traffic through priority scheduling information \cite{CSR5}. 
However. the PSCR has been thoroughly researched and studied in cellular networks. For instance, a mathematical model for joint and fair transmission scheduling and a power spectrum adaption algorithm for CSR was proposed in a study referenced as \cite{CSR6}. The study confirmed that PCSR approaches can significantly improve network throughput. Another study referenced as \cite{CSR7} presented the performance analysis of PSCR and coordinated scheduling to utilize spatial reuse over cellular networks. Additionally, several works have discussed coordinated scheduling methods for CSR in cellular networks, including references \cite{CSRN1, CSRN2, CSRN3, CSRN4}. While PCSR has been extensively studied in cellular networks, there is a lack of research on its implementation in WLAN. However, there are a few articles, such as \cite{pscr-wlan}, that offer guidance on implementing PCSR using a Q-learning-based approach to share information about scheduled APs. The proposed approach aims to avoid sharing information about interference-free APs by comparing Q values. Additionally, \cite{kihira} proposes an adversarial reinforcement learning-based CSR method to reduce frame signaling overhead with partial receiver awareness, which could be beneficial for highly dense WLANs. Moreover, ~\cite{chem} proposed a scheduling approach to enhance the efficiency of PCSR for real-time applications (RTAs) within Wi-Fi 8 networks without compromising the performance of regular STAs. It classifies regular STAs into PCSR-favorable and PCSR-unfavorable groups based on their ability to support reliable RTA transmissions. PCSR-favorable STAs meet a minimum SINR threshold, while PCSR-unfavorable STAs cannot meet the condition or disallow PCSR during transmission. The success relies on coordinated efforts between the regular AP and RTA AP, where the regular AP shares interference information, and the RTA AP reports received power levels. This strategy ensures efficient PCSR utilization for RTAs while maintaining regular STA functionality, emphasizing collaborative coordination and measurement processes.

\textit{MAC-based CSR}: The MCSR addresses various concerns including determining AP-STA pairs for CSR, grouping STAs that are free from interference, and implementing CSR based on graph coloring. In an example described in \cite{CSR8}, multiple APs within range are connected over the air to share control information. A MA is selected to win the TXOP first, while other nearby APs act as slave/coordinated APs. The proposed approach forms groups with similar interference levels and determines TPC parameters for each group type. A reasonably distant AP is chosen to form a CSR group, restricting the use of immediate neighboring APs and STAs at the edge of neighboring APs. However, this approach only considers static STAs and WLAN environments, which may not be effective in dynamic WLAN environments with new AP deployments and moving STAs. In \cite{CSR9}, the authors proposed TXOP sharing methods for CSR over 802.11be WLAN, using mathematical models to determine mutual interference and choosing combinations that provide minimal interference and higher throughput. However, this article lacks analysis on the interval of finding combinations, and the solutions have high complexity issues in dense AP environments like mmWave WLAN. As finding the number of combinations using brute force is an NP-hard problem, the use of evolutionary algorithms may be useful. Another MAC-based CSR method addresses the challenge of simultaneous multi-AP transmissions, aiming to maximize spatial reuse opportunities for optimized resource utilization in the network~\cite{david}. After establishing multi-AP groups, the study delves into various scheduling algorithms for MAP-Co transmissions. Two categories of algorithms are examined: per-AP algorithms, relying on individual APs' buffer states, and per-group algorithms, considering the collective buffer state of all APs within a group. Notably, per-AP algorithms outperform per-group counterparts, minimizing worst-case delay by prioritizing APs with higher packet counts or older waiting packets in the buffer. This approach significantly reduces data transmission time, enhancing overall network performance. The authors acknowledge the study's focus on a small network with nine APs and a maximum of three APs per group, highlighting the need for further analysis under conditions of a larger network with more APs per group.


\subsubsection{Summary and Lesson Learned}
This section covers the difference between CSR and SR, as well as the architecture, preparation stage, and data transmission process of CSR. We discovered that in the CSR architecture, it's advisable not to fix the coordinator in the MA-based architecture to allow for flexibility and avoid burden on one AP. We also provided formulae to calculate the transmission power for an AP that will have SR in CSR, based on channel sounding information from multiple APs. According to equation.\ref{csr-eq}, the allowed interference level should be greater than the RSSI values from neighboring APs. These mathematical formulae can help mathematically model complex situations of CSR for TPC.  We have categorized CSR methods based on when they are performed: Periodic and Every-Transmission. After discussing these two types, we have found that periodic transmission has lower overhead and energy consumption compared to Every-Transmission, which requires more frames to be exchanged. However, periodic transmission may not be suitable for dynamic WLAN environments like future extreme high-frequency bands (emerging 802.11ay and Terahertz(THz)), which experience frequent changes. On the other hand, Every-Transmission based CSR may not be suitable for the emerging 802.11be WLAN due to its significant overhead. As a result, a new CSR transmission method needs to be developed in the future.

Additionally, we have showcased the current state-of-the-art in CSR methodologies, which are conveniently summarized in Table~\ref{table-csr}. After conducting thorough research on existing CSR literature, we have introduced a classification system that divides CSR approaches into three categories based on their optimization techniques: OPSCR, PCSR, and MCSR. OPCSR approaches employ various methods to optimize the transmission power of OBSSs in a coordinated manner, thereby minimizing mutual interference during concurrent transmission. In Table ~\ref{table-csr}, we can observe that the majority of OPCSR techniques are designed for DL transmission, which goes from the AP to the STAs. It is difficult to implement OPCSR for UL transmission because synchronization is time-consuming and costly. Additionally, we can infer that OPCSR can be either periodic or applied to every transmission. In Table.~\ref{table-csr}, we have listed the downsides of each OPCSR method discussed. However, the PSCR approach is versatile and can be used for both UL and DL transmission, as shown in Table.~\ref{table-csr}. Additionally, PCSR has several advantages over OPCSR, including reducing contention time during CSR, allowing more concurrent TX through high resource utilization, and reducing latency for time-sensitive traffic through priority scheduling. Therefore, PCSR is recommended for both UL and DL transmission in emerging WLAN. Nevertheless, implementing PCSR may present some challenges. When it comes to UL transmission in WLAN, STAs won't know if there's any interference at APs because there's no feedback system in place. Additionally, some STAs have to wait for other STAs to finish transmitting in CSR before they can start. Unfortunately, there isn't a mathematical model to figure out the average delay STAs experience during UL transmission using PCSR. That's where MCSR comes in, which helps identify interference-free APs and STAs pairs and groups them for simultaneous transmission. By using the grouping approach and identifying interference-free AP-STA pairs, the overhead can be reduced. It's important to redesign the MAC layer of emerging WLANs to include a CSR-based grouping of STAs.\newline
Most CSR approaches rely on predetermined frequency-time allocations by coordinators. This limits their ability to quickly adapt to changes in user distribution, QoS requirements, and channel conditions. Additionally, these approaches are reactive rather than proactive, meaning CSR transmission and decision-making occur after several rounds of channel information gathering and analysis. This reactive approach can cause delays that are not suitable for URLLC applications, especially in highly dense WLANs like mmWave WLAN. Therefore, there is a need for a proactive CSR approach that can predict future CSR transmission and determine scheduling details in advance. Currently, CSR approaches have been well studied for 802.11be WLAN but have not explored the challenges specific to mmWave WLAN (802.11ay). When there are many active users simultaneously, maintaining orthogonality among users becomes difficult, leading to interference from various sources. Therefore, it is essential to identify scenarios where mutual cell interference can occur and propose new management policies for CSR.


 \begin{figure}[t]
	\centering
	\includegraphics[width=3.5in]{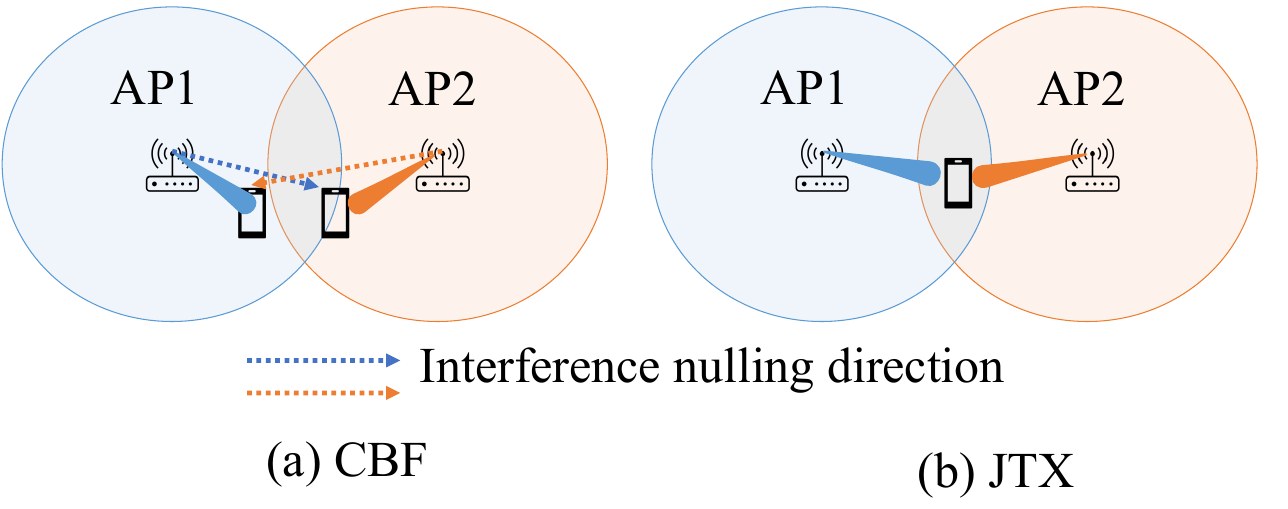}
	\caption{Basic concept of CBF and JTX.}
	\label{cbffig}
	\vspace{-0.5cm}
\end{figure}

    \subsection{Coordinated Beamforming}
The CBF concept aims to improve spatial reuse using same transmission power by enabling cooperative APs to cancel incoming interference at the spatial level, as shown in Fig.\ref{cbffig} (a). For instance, if there are two coordinated APs, each with an STA, both STAs can transmit data to their respective APs using spatial reuse TXOP. In CBF, the APs work together to share information on the interference their STAs receive, which can be used for interference nulling or alignment. Non-serving STAs can provide CSI by using MAP-Co channel sounding methods to help APs understand the interference they receive. CBF is a popular topic in cellular networks and has recently been discussed in unlicensed WLANs, such as 802.11be~\cite{cbflte1, cbfwlan}. Recent results show that MIMO antennas can suppress interference, reducing neighboring link interference by up to 10 dB~\cite{cbflte2}. This section provides a detailed explanation of CBF operations and the existing state-of-the-art. 
\subsubsection{CBF Operations}
The process of CBF operations is similar to the previous MAP-Co features and consists of three phases: MAP-Co request, channel sounding, and MAP-Co data transmission. The MA/MC sends MAP-Co requests to its STAs and slave APs, which then forward the request to their respective STAs. Afterward, channel sounding is performed, which is explained in detail in previous sections. In this section, we will focus on the MAP-Co CBF data transmission steps. 
We consider a WLAN network with two OBSS that have two APs (AP1 and AP2), each having an STA associated with it (STA1 to AP1 and STA2 to AP2). AP1 and AP2 are slave APs, and AP1 needs to transmit DL to STA1 while STA2 needs to send UL packets to AP2. However, AP1 DL transmission can cause interference at STA2, while UL transmission by STA2 can cause interference at AP1. To avoid interference, the MA/MC sends a trigger frame to slave APs that conveys synchronization, scheduling, and interference nulling information. AP1 and STA2 synchronize their data transmissions within the same TXOP of AP2 and STA2 and utilize coordinated sounding information to optimize the antenna weights of their respective antenna arrays. The goal is to suppress and nullify interference at STA1 by adjusting the antenna weights. In the next section, we will present the state-of-the-art of interference management through CBF.

    \subsubsection{State-of-the-art of CBF}
 In the context of CBF, the primary objective is to effectively handle interference in a multi-AP environment. This implies that when an AP transmits data to an STA, it should not disrupt the reception of another STA in the adjacent OBSS. Interference management in CBF can be classified into two main categories: PHY layer-based and MAC layer-based approaches. The MAC layer-based approach can be further subdivided into centralized and semi-distributed schemes. Similarly, the PHY layer-based approach can be categorized into two types: Interference Nulling and Interference Alignment.

In centralized schemes, an MC-based architecture can be employed to coordinate interference management through beamforming among multiple slave APs. On the other hand, semi-distributed approaches utilize MA-based architectures to manage interference. Interference nulling is a technique used by a transmitting AP to completely cancel its signal at a specific receiver by precoding the signal to counteract the interference caused by the signal transmitted to another receiver. In contrast, interference alignment aligns all interfering signals in the same directions at the receiver, allowing the signal from a particular transmitting AP to be free from interference and enabling successful decoding. For more in-depth information regarding interference nulling, refer to \cite{null1}, while details on interference alignment can be found in \cite{null2}. Now, we will delve into the existing literature pertaining to each category of interference management in CBF.

  One of the previous studies, identified as \cite{cbf1}, concentrated on managing interference in a centralized manner for a multi-AP setting that is coordinated using a backbone network. In this method, all the slave APs shared data packets through a wired backbone network, and the precoding process was similar to MU-MIMO transmission. The method employed interference nulling in CBF and required explicit channel sounding prior to CBF. However, the drawback of this approach is the overhead introduced by explicit channel sounding, which increases the overall overhead and renders it unsuitable for densely populated real networks.
    Therefore, the approach proposed by the authors in \cite{cbf2} uses implicit channel sounding and offers higher accuracy. These papers are excellent examples of comparing the CBF approach using different channel sounding methods in a centralized way. Another method, discussed in \cite{cbf3}, focuses on a centralized interference alignment approach for CBF. In this approach, the centralized server performs joint precoding of transmitted signals to APs and transfers the frequency domain symbols, ensuring that all slave APs are phase synchronized. The method utilizes low complexity zero-forcing beamforming and Tomlinson-Harashima precoding.
In \cite{cbf4}, the authors put forth a centralized interference alignment technique known as POLYPHONY. Essentially, their method involves directing the signal from a station towards all access points (APs) in the same direction to maximize the SNR for packet decoding. This process is repeated for every packet from each station. While this approach is reliable, it can create significant overhead by sending the same packet to each AP, and it only applies to the uplink (UL).
   The authors of \cite{cbf5} suggested a centralized method for interference alignment and nulling in both the uplink and downlink using MIMO technology. Their approach involved using an antenna to nullify a signal received by an STA, and then transmitting another signal to another STA through its antenna to align with the received signal.
  A proposed approach by \cite{cbf6} also uses a centralized method that combines signal processing to create a joint precoding matrix in the downlink and joint decoding in the uplink transmission for interference alignment. This approach also uses explicit channel sounding to gather CSI information. An experimental study confirmed that this method offers optimal multiplexing gain. Other works, such as \cite{cbf7}, also focus on joint precoding and decoding for interference alignment.
  
 Another concept that focuses on CBF is cell-free networks, which introduce a new way of utilizing antennas in a coordinated manner~\cite{cellfree1}. This means that the principles of cell-free networks and CBF techniques can be integrated into MAP-Co, and advanced physical layer concepts from cell-free networks can enhance MAP-Co performance.
One of the primary research areas in cell-free networks is determining precoding matrices for transmitting signals from multiple antennas~\cite{cellfree2,cellfree3}. This can reduce destructive interference at the receiver. Conjugate Beamforming precoding is a commonly used technique to improve wireless communication in these networks~\cite{cellfree4}. CB precoding involves manipulating the phase and amplitude of transmitted signals to maximize the desired signal's strength while minimizing interference and power consumption.
However, cell-free networks lack channel hardening characteristics, which can increase self-interference at the receivers~\cite{cellfree3}. To address this issue, researchers have proposed a modified conjugate beamforming approach. They suggested a mathematical model to update the precoding matrix, which scales down the precoder if the gain of transmitting APs to a user exceeds a target value bringing the gain to the target value~\cite{cellfree5}.

Another approach for precoding in cell-free networks is the centralized zero-forcing (ZF) method, where the ZF coding method is applied for multiple distributed antennas to decide precoding matrices centrally, to force transmitted signals from multiple antennas to be orthogonal concerning interference from other users~\cite{cellfree6}. However, the ZF approach for large-scale antennae is computationally expensive. Hence, most of the works related to precoding in cell-free networks have focused on the design of cost-effective advanced precoding methods that are computationally efficient.

One proposed approach is the local full-pilot zero forcing (FZF)~\cite{cellfree7}. The core concept is to eliminate the interference caused by co-pilot users effectively. However, it is important to note that its effectiveness depends on having more antennas at the AP compared to the length of the pilot sequence. If this condition is not met, the advantages of FZF precoding are reduced as the number of antennas at the APs decreases.
Therefore, researchers have proposed Local Partial Zero-Forcing (PZF)~\cite{cellfree9}. PZF is a wireless communication strategy employed at access points (APs) to enhance data transmission efficiency. It classifies users into two groups: "strong users" and "weak users," based on their channel quality using predefined threshold values. Strong users receive data using FZF precoding, optimizing their data transmission. Weak users, however, use conjugate precoding, which directs signals to minimize interference from strong users. Local PZF relaxes the strict FZF requirement and offers an extra layer of interference protection for weak users by ensuring that their conjugate precoding takes place in orthogonal channels relative to strong users. This approach tailors precoding techniques to individual users' channel conditions, ultimately improving system performance.
The results revealed that the PZF approach provides a higher data rate compared to FZF. In this regard, researchers compared different precoding methods in a cell-free environment~\cite{cellfree9}. They considered 100 APs, each equipped with ten antennas to serve 20 users with a pilot signal sequence of length 5. Through their analysis, they proved that the conjugate method has the lowest achievable data rate compared to centralized ZF, FZF, and PZF. Centralized ZF provides a 12 bps/Hz data rate with a cumulative distribution probability value of 1.
Another approach to decide precoding matrix for CBF is the use of matrix inversion approach. The matrix inversion approach aims to maximize the SINR at the receivers. It does this by directly inverting the channel matrix to obtain the precoding matrix. For instance, in a study, the authors of \cite{cellfree10} proposed a model using different iterative matrix inversion methods in cell-free networks. They identified that hermitian-preserving methods are applicable for cell-free networks due to their support for parallel computation, low latency, and numerical stability. These methods can be used to minimize interference during linear precoding, which can be computationally intensive. Therefore, it is crucial to find efficient ways to perform matrix inversion to reduce the computational complexity of precoding algorithms.
Such advanced precoding approaches based on cell-free concepts can also be applied to the MAP-Co environment.
  
   Some approaches prioritize CBF over MA-based architecture to minimize the burden on a central server. An example is the CoaCo method proposed by the authors of \cite{cbf8}, which uses semi-distributed CBF to address interference in WLAN. Their method optimizes beamforming weights to reduce inter-AP interference through grouping, ensuring that interference between groups is minimized. Coordination is only required between the heads of each group, similar to MA-based architecture.
    Another method called OpenRF has been proposed for MA-based architecture to nullify interference. This approach uses software-defined networking (SDN) to target the beam to a specific STA and automatically nullify any interference at other STAs. The implementation of SDN-based interference alignment has also allowed for an increase in concurrent data streams.
    In \cite{cbf10}, the authors suggest a cluster-based CBF for MA-based architecture. This method utilizes decentralized CSMA to prevent interference between clusters through a precoding method within the cluster. Other research has also addressed inter-cell interference during beamforming through clustering methods, such as \cite{cbf11,cbf12,cbf13}.

    \subsubsection{Summary and Lesson Learned}
In this section, we explored different methods for improving concurrent transmission using CBF. Centralized and semi-distributed interference nulling and alignment approaches were discussed. Centralized schemes rely on the network's backbone capacity for efficient information sharing, making them suitable for enterprise WLAN but not ideal for home-based WLANs that require low latency and high bandwidth. Interference nulling and alignment have been well studied for single user MIMO and MU-MIMO, but not enough attention has been given to dense multiple AP scenarios. Precise CSI information is needed for precoding, and existing reactive CSI gathering schemes may not meet future extreme requirements.

We also looked at a clustering-based approach for interference management in beamforming. STA selection is crucial within the cluster for beamforming scheduling. However, existing methods use a greedy approach for user selection. In the future, it would be noteworthy to study the design of an evolutionary algorithm for user selection in large-scale networks. Semi-distributed approaches are promising for addressing scalability issues related to CBF.

  \subsection{Joint Transmission}
We previously discussed the efficient management of data transmission by multiple STAs of OBSSs using techniques like C-OFDMA, CSR, and CBF. In addition to these features, MAP-Co also supports efficient distributed MIMO (D-MIMO), which enables transferring packets between multiple APs and an STA. This concept has been around for a decade and has been implemented over cellular networks. However, researchers are now considering D-MIMO over emerging WLAN technologies like 802.11be~\cite{deng}. D-MIMO is a significant feature of MAP-Co and is also known as JTX. As depicted in Fig.\ref{cbffig} (b), in JTX, multiple slave APs can transmit data to an STA by receiving JTX scheduling and control information from MC/MA.
JTX offers numerous advantages, including seamless association for mobile stations (STAs), increased throughput, improved reliability, and reduced transmission delays. One example of JTX operations in the context of WLAN is its ability to avoid ICI by enabling multiple overlapping basic service sets (OBSSs) to transmit on the same channel simultaneously. In this section, we will discuss an example of JTX operations specifically in the context of 802.11be and provide an overview of the current state-of-the-art in JTX. Finally, we will summarize the key findings and lessons learned from the discussion.
   \subsubsection{JTX Operations}   
In this section, we will discuss the process of performing JTX from the perspective of both MAP-Co architectures. Prior to initiating JTX operations, it is important to conduct multi-AP channel sounding, which involves characterizing the wireless channel environment at multiple access points (APs). In section~\ref{channelsounding}, we explained various multi-AP channel sounding methods. However, all of those methods were focused on single AP-STA transmission. Therefore, a new method called JTX channel sounding was proposed by IEEE for implementing the same process in JTX. During JTX channel sounding, the coordinator (MA/MC) initiates the process by sending an NDPA frame to all slave APs involved in JTX. This frame contains information about the intended STA for JTX and instructs the directly associated AP to send an NDP trigger frame to intended STAs, which broadcasts NDP packets from the STA. The other slave APs are instructed to be ready to receive the NDP frames from the STA. Once all slave APs receive the NDP frames, they estimate the CSI and report it back to the coordinator. Only the slave APs involved in JTX participate in this process.

Next, the coordinator analyzes the reported CSI and sends a BRP trigger frame to the selected slave APs for JTX. However, if a slave AP is busy and unable to participate in JTX, the BRP frame is not transmitted to it. The BRP frame contains scheduling and control information for JTX, including power levels, phase shifts, and data frames. To ensure synchronized transmission to the intended STAs, control parameters are used to strengthen the transmitter.

Upon receiving the BRP trigger frame, the slave APs initiate joint data transmission to the STA. The STA sends ACK packets to the slave APs, and the coordinator requests ACK acknowledgments from the slave APs. When the request is received, the slave APs forward the ACK acknowledgments. JTX manages tight synchronization among the slave APs to ensure that data transmission starts and ends simultaneously.

In IEEE 802.11, multiple access by STAs in JTX can be achieved using C-OFDMA and CBF. Slave APs can utilize different primary or secondary channels through C-OFDMA for DL JTX to an STA or employ CBF to transmit in different sectors to an STA. During JTX, the slave APs configure their transmission parameters to ensure constructive and joint decoding of the signal at the STA, resulting in a higher SINR. Similarly, CBF and C-OFDMA can be applied for UL JTX by multiple STAs to an AP, and joint processing, such as joint decoding, is performed at the AP. The STA needs to perform joint processing of the received signals for effective decoding.

    \subsubsection{State-of-the-art of JTX}
    In the previous section, we covered various steps involved in JTX operations over WLAN. Now, we will introduce the latest advancements in JTX through a new classification system. The state-of-the-art of JTX can be categorized into three types, each addressing a distinct problem: Joint Scheduling, Synchronization, and Joint processing. 
    

    \textit{JTX Scheduling:}

    In JTX, one important aspect is deciding on the intended STAs and scheduling policies for resources. This process typically involves three steps: selecting the transmission points, pairing them with the corresponding STAs or users, and determining resource allocation parameters such as power levels and MCS.  Researchers have explored various methods for joint scheduling approaches, which is also a well-researched topic in cellular networks. To begin our discussion, let's examine joint scheduling in cellular networks and its limitations in WLAN. 
    
   The authors of~\cite{JT1} proposed a centralized grouping-based MAC scheduling approach for multi-point JTX. In this approach, the base stations are grouped in different clusters with different strategies while considering fixed transmission points and groups. The approach is a good study of different strategies, but it has a few limitations. Firstly, it is not applicable for more than three base stations in the group, and secondly, it is not suitable for mobile STAs. Furthermore, the delay can be very high in case of larger access points, such as in the case of WLAN. Another disadvantage of the static cluster is the consideration of fixed interference, which is not practical in the case of WLAN. Therefore, such grouping approaches cannot be applied to the JTX over WLAN. 
    In a study cited as \cite{JT2}, the authors suggest a clustering-based multi-point JTX approach and allocate resources to users within a cluster through Space-division multiple access. However, the authors limit the number of users who can perform JTX to the number of cells in the cluster. This restriction can cause significant delays in WLAN due to the CSMA/CA process and the higher number of APs. Additionally, the approach assumes static devices and fixed groups, which may not be practical for dynamic changes in the WLAN environment, such as interference and mobile devices. While this method is applicable for WLANs that consider interference and power levels, it is too basic and does not consider WLAN channel access models. Therefore, determining JTX and non-JTX users using this straightforward approach may not be suitable for highly dense WLANs, leading to longer delays due to the CSMA/CA process. Other works have considered dynamic clustering methods for JTX.
   For an example, an user-centric approach based on location was proposed by \cite{JT5} for creating small virtual cells for a JTX to a user using multiple access points. The approach considers power constraints for each user and uses mathematical models to estimate average user throughput. This method is suitable for 802.11be WLAN due to its wider coverage. However, clustering based on user location may result in resource wastage if available bandwidth does not meet the required QoS. Therefore, user location alone may not be the most efficient approach for JTX.
    Another method for clustering APs in JTX involves selecting a certain number of users and scheduling resources based on interference and the optimal power level difference between the base station and the users~\cite{JT4}.

   In addition to clustering, another method for determining the number of APs needed for JTX is a graph-based approach outlined in \cite{JT3}. The authors proposed an OFDMA-based approach for JTX in cellular networks, where they created graphs of base stations involved in JTX, with each node assigned its queue size and traffic arrival. By using graph coloring and the knapsack problem, they were able to solve the JTX resource scheduling issue and improve the throughput for inter-cell users. In their  \cite{JT4}, the authors proposed a novel method for choosing user-AP pairs in JTX that do not rely on clustering or graphing. Instead, they suggest determining the optimal power level difference required for a certain number of users to operate in JTX mode under an AP. They also recommend radio resource management for JTX users, which can be beneficial for smaller networks that prioritize individual users over APs. However, this method may not be practical for denser networks due to synchronization requirements and network delay. Additionally, scheduling approaches for JTX cannot be directly applied to WLANs because of differences in multiple access methods, path loss, blockage, interference, and QoS requirements. In addition, WLANs typically lack X2 interfaces which are needed for synchronization and sharing. As a result, there have been very few studies on JTX conducted over WLANs. We will now showcase the most recent developments regarding JTX over WLAN.
   
In 2018, a group of researchers proposed a JTX method specifically for 60 GHz WLANs~\cite{JT6}. Their method accounted for link outages caused by human body blockages and introduced a blockage mitigation technique using JTX. This technique ensured reliable communication even when blockages were present. The researchers carefully selected STAs for JTX to improve reliability and capacity. They also addressed interference issues by determining the optimal JTX path, taking into consideration the presence of potential sources of interference. However, the researchers had limited control over available paths, which could result in interference problems, especially when dealing with a higher number of mmWave APs.
In order to enable JTX in crowded places where interference-free paths are not available, the authors of \cite{JT7} suggested using an IRS based JTX to create new paths. Additionally, there is a channel reservation based MAC protocol for JTX in WLAN, introduced in \cite{JT8}, which can enhance the throughput and reliability of a STA.
A method for analyzing the performance of JTX over 802.11be, as well as a distributed scheduling scheme for JTX, single AP, and single AP transmission with interference, was proposed by \cite{JT9}. The goal of their work is to allow JTX for an STA that is under more than one AP. However, this approach may activate JTX on multiple STAs that do not require it, leading to reduced resource utilization. As a result, more research is needed to address WLAN challenges and advance JTX scheduling. 
\begin{figure*}
	\centering
	\includegraphics[width=7in]{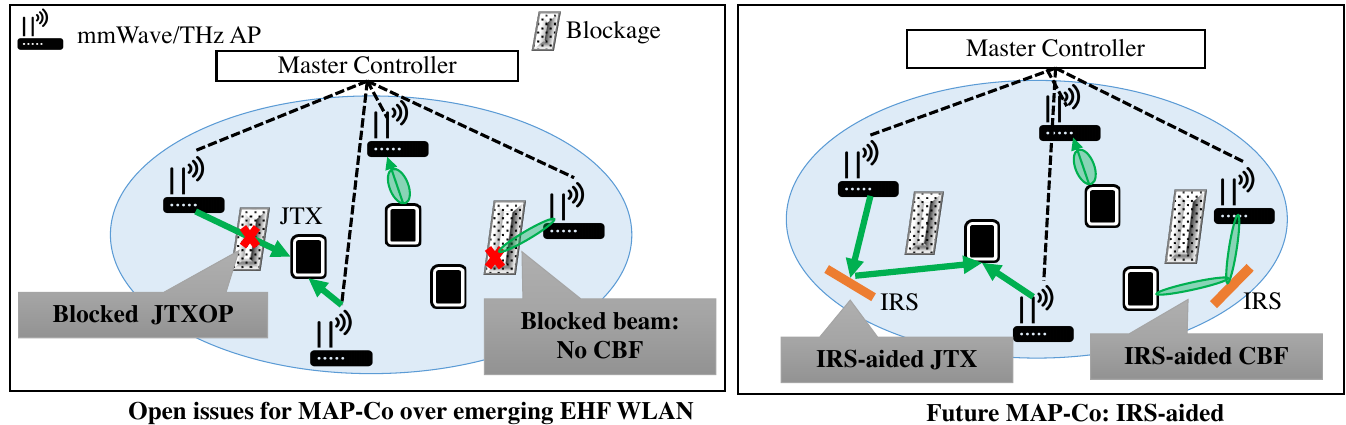}
	\caption{Blockage-Aware MAP-Co: Open issues and future direction.}
	\label{future1}
\end{figure*}
\newline
\newline
\textit{Joint Synchronization and Processing:}

To ensure smooth operation of the JTX, it's important to synchronize the joint process from multiple APs. This will prevent any timing offset, allow for coordinated phase shift, and synchronize the digital clock at the STAs. Failure to synchronize can result in inaccurate channel value estimates, poor interference nulling, and degraded SINR at the STAs, as noted in \cite{JT10}.
Moreover, in JTX scenarios, accurately estimating CSI by each STA becomes increasingly difficult over time. This creates a challenge for synchronization. To address this challenge, AirSync \cite{JT11} proposes a method for timing and phase synchronization. This method detects slot boundaries in OFDM, synchronizes APs with cyclic prefix, and predicts carrier phase correction for transmitters. While this method is suitable for 802.11ay WLAN, it cannot be applied to 802.11be WLAN due to differences in channel modulation techniques.
Joint processing between AP and STAs is also crucial, encompassing tasks such as estimating joint CSI, performing joint precoding, and handling backhaul processing. This allows for enhanced desired signals, overpowering any interference or noise, by combining signals from multiple APs at an STA. To achieve this, all signals need to undergo joint processing. The maximum ratio method is one way to merge received signals, optimizing the ratio between the power of the desired signal and the squared norm of the combining vector \cite{JT12}. However, this technique is not optimal for dealing with interfering signals. Another approach involves a tradeoff between interference rejection and maximizing signal gain \cite{JT13}. In \cite{JT14}, various other methods for joint processing of reference signals are surveyed. Furthermore, joint processing of transmit precoding is essential for decoding joint reception at receptors, and one approach involves precoding through uplink-downlink duality using linear combination schemes for joint precoding vectors \cite{JT15}.

    \subsubsection{Summary and Lesson Learned}
In our previous discussion, we talked about the JTX process, where an STA receives data from multiple APs, also known as transmission points. In addition, we have discussed the JTX operations recommended by the 802.11be standards for JTX over WLAN. We also provided an overview of the current state-of-the-art JTX scheduling methods. Many of these methods use a grouping and clustering approach to organize the APs involved in JTX for an STA and allocate resources based on their needs. However, this approach has lower scalability and is best suited for static WLAN environments with a fixed number of STAs and APs. In case of dynamic and dense mmWave WLAN environments, this approach may cause delays and reduce reliability. A dynamic clustering approach, such as the one presented in \cite{JT4}, could be a suitable option for future WLANs. However, it's important to note that the efficiency of JTX may decrease in the presence of obstacles in mmWave WLANs. To address this issue, blockage-aware solutions, as discussed in \cite{JT7}, should be studied. Another challenge with JTX is joint synchronization and processing. Improper synchronization can result in a different MIMO channel between APs and STA, causing poor JTX. JTX scheduling should also consider channel aggregation/bonding, which hasn't been fully explored in the literature. Additionally, joint precoding and decoding of massive JTX in future dense mmWave WLANs can create overhead and longer delays. Therefore, an intelligent and proactive JTX solution is necessary to adapt to the ever-changing WLAN environment, with a focus on designing distributed precoding and decoding processing of JTX.

\begin{figure*}
	\centering
	\includegraphics[width=7in]{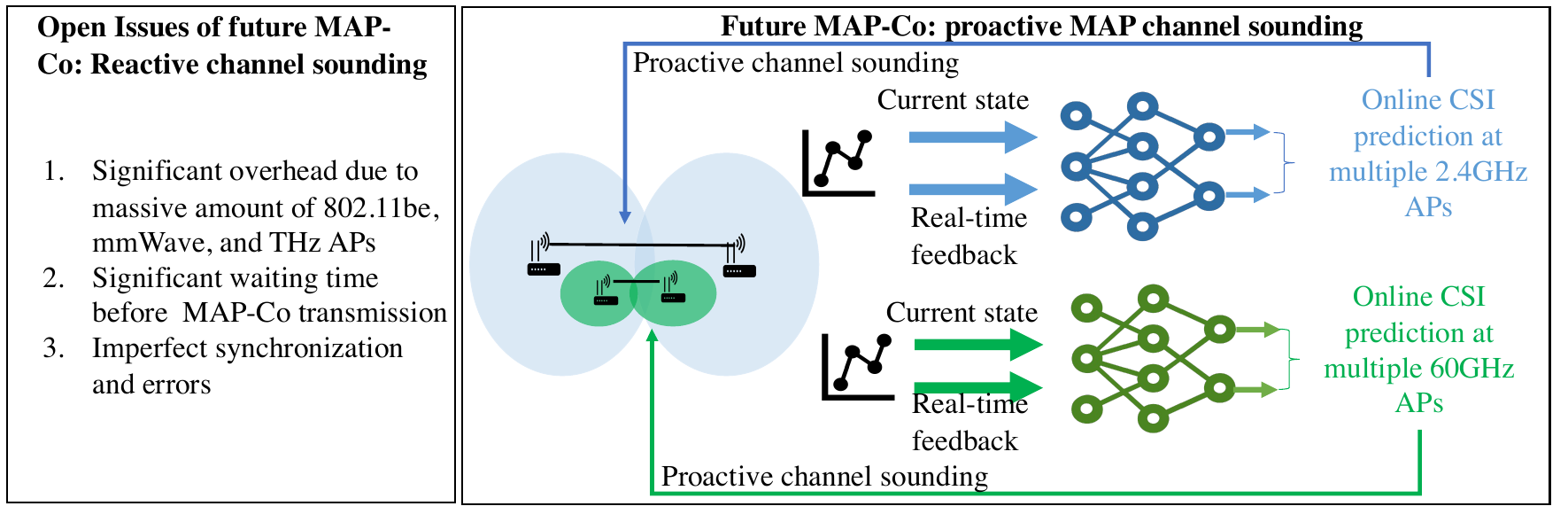}
	\caption{Proactive channel sounding: Open issues and future direction.}
	\label{future3}
\end{figure*}
\begin{figure*}
	\centering
	\includegraphics[width=7in]{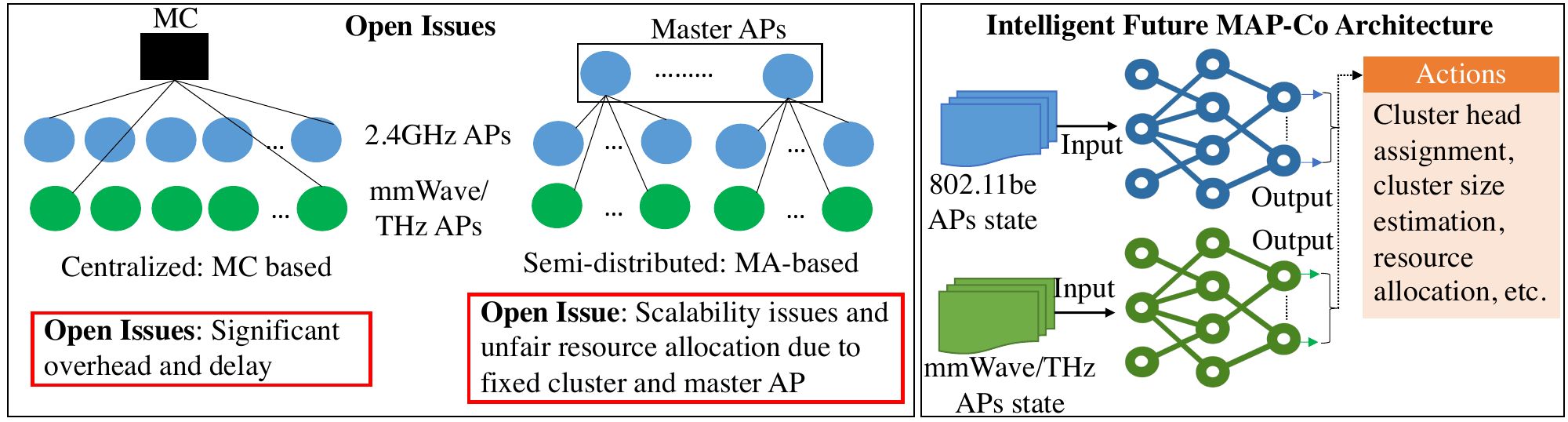}
	\caption{Future MAP-Co architecture: Open issues and future direction.}
	\label{future5}
\end{figure*}
 \section{Future MAP-Co and Challenges}\label{futuresec} 
 In this section, we present several promising future research areas for MAP-Co and open technical challenges for the realization of future MAP-Co. The explained future areas and challenges are going to promote the development of MAP-Co in the emerging WLANs such as beyond 802.11ay, beyond 802.11be, and THz. The details of each future area of MAP-Co are explained in the next subsections. 
 \subsection{Blockage-Aware MAP-Co}
Future MAP-Co will be implemented over EHF bands such as mmWave or THz. However, the EHF bands are highly vulnerable to WLAN environments such as obstacles or humans, which can degrade signal quality and strength. Unfortunately, existing MAP-Co features such as channel sounding, CBF, CSR, and JTX do not consider the impact of blockage on their performance.
For instance, the existing CBF does not take blockage-aware coordination into account. As a result, beams from a blocked AP cannot perform transmission at the same TXOP with other APs, and cannot take advantage of CBF. Similarly, if power control in CSR is oblivious to blockages, then the estimated power may not be enough for a high-quality signal. Therefore, transmission power control in CSR should consider the blockage between STA and AP.
Blockage in MAP-Co over mmWave WLAN can also impact the performance of JTX. For example, JTX by multiple APs to STAs can be affected if there is a blockage between the STA and a particular AP, as presented in Fig.~\ref{future1}. In such cases, the transmission by that AP can reach the STA with very low quality~\cite{niyato2}, reducing the overall performance of JTX. Moreover, existing models for calibration in implicit channel sounding do not consider the impact of blockage, leading to inaccurate CSI estimates. Therefore, blockage-aware MAP-Co design is another open issue.

One solution to these issues is to use an intelligent reflecting surface (IRS) to improve the propagation environment by converting non-line-of-sight to line-of-sight without signal loss~\cite{niyato1}. An IRS can also expand the coverage of JTX and other MAP-Co features, as depicted in Fig.~\ref{future1}. However, using an IRS for MAP-Co performance poses several challenges that need to be addressed. Researchers need to propose new mathematical models to understand multiple channels in IRS-enabled MAP-Co, and take into account the extra signal processing required at the master AP. Moreover, the placement of the IRS in the context of MAP-Co is a key investigation to avoid interference at STAs or other APs from the reflected signal. Another direction could be the design of new calibration models for implicit sounding that consider the signal losses due to blockage. Therefore, in the future, MAP-Co shall be IRS-assisted.

\subsection{Proactive MAP-Co Channel Sounding}
  After studying various MAP-Co channel sounding methods, we have identified two prevalent problems. Firstly, there is considerable signal overhead when using explicit channel sounding. Secondly, implicit channel sounding can result in calibration errors, as indicated in Fig.~\ref{future3}. These issues arise due to the reactive nature of multi-AP channel sounding, which involves sending reference signals or null packets. Hence, an intelligent channel sounding is a promising approach to address the issues of signal overhead and calibration errors in multi-AP channel sounding, as depicted in Fig.~\ref{future3}. Various methods have been developed to predict CSI by examining a user's past mobility patterns or handover patterns to forecast future locations and estimate the CSI from that location~\cite{aicsi1,aicsi2,aicsi3,aicsi4,tiago}. Other approaches analyze an STA's past CSI information within an AP to predict the CSI. Many of these methods employ deep convolutional neural networks (CNN) to forecast the future CSI that an STA will experience at an AP. However, existing deep learning-based approaches for single-AP CSI prediction cannot be directly applied to multi-AP environments due to the increased complexity of input and output parameters. Therefore, developing new approaches that can reduce the scale of input and output parameters is an open area of investigation. 
Hence, we propose a novel multi-AP deep-CNN model idea for CSI prediction for proactive MAP-Co channel sounding. In the multi-AP deep-CNN based CSI prediction, the past CSI information of STAs at multiple APs at each noted time should be considered. Additionally, the output of the multi-AP deep-CNN model should predict the CSI at multiple APs at each noted time. However, the scale of the output will be too large to train the multi-AP deep-CNN model. Therefore, an approach needs to be implemented to optimize the multi-AP deep-CNN model. 
As a result, we present an idea to optimize the multi-AP deep-CNN approach. Our idea is to predict the CSI of a STA at multiple APs by predicting the time interval when STAs are at the edge of OBSS. The multi-AP channel sounding approach mostly benefits the edge STAs by avoiding interference and improving network performance. Since an STA stays at the edge for shorter time intervals, estimating CSI from all slave APs at each interval is a waste of resources and increases the complexity. Furthermore, the weight values and other neural network parameters need to be adjusted accordingly.
  
  In addition, implementing an intelligent channel sounding method is simple with MC-based MAP-Co. However, a centralized multi-AP deep learning approach is not feasible for MA-based MAP-Co due to its higher computational demands. Federated learning can be a promising approach for implementing intelligent channel sounding in MAP-based MAP-Co. Federated learning is an approach to improve network by using decentralized learning at multiple local servers~\cite{niyato3,kato1}. In this approach, local servers at each slave AP can predict the channel quality of STAs locally, and the predicted results can be sent to the AP for multi-AP prediction. However, due to the highly dynamic nature of the mmWave WLAN environment, the accuracy rate of the prediction can decrease with even small changes in the environment. Therefore, the multi-AP deep CNN models for 802.11ay should be designed to ensure that changes in the WLAN environment do not reduce the accuracy rate for an extended period of time. This can be achieved through continuous monitoring and re-training of the deep learning models, which would help maintain a high accuracy rate for multi-AP channel state prediction in dynamic WLAN.

\subsection{Multi-band MAP-Co}
Wireless Local Area Networks (WLAN) operating on Sub-6 GHz bands may not be sufficient for advanced applications like virtual reality (VR) and may not provide seamless connectivity. To address this issue, multiple bands such as 2.4 GHz/sub-6 GHz and mmWave are being used in WLAN development to ensure seamless connectivity and meet extreme requirements~\cite{multi-band,zubair2}. Moreover, The upcoming 6G wireless network aims to integrate a diverse range of networks, including cellular and WLANs, that cover bands ranging from sub-6 GHz to THz. This feature will result in small and highly dense cells. However, coordinating existing methods across different bands is challenging due to different transmission protocols. This makes it tough to effectively manage the network and improve the quality of service (QoS) in 6G. To overcome the challenges faced by WLAN in dense scenarios, \cite{pzhou} proposes a centralized control architecture that separates the control (sub-6 GHz) and data planes (mmWave). This approach optimizes radio resource allocation and reduces mmWave beamforming training overhead. The joint resource allocation of control and data planes significantly improves user scheduling and overall sum rate in simulations, surpassing traditional methods.

Apart from integrated heterogeneous multi-band networks, the 802.11 protocols also offer a way to quickly transfer sessions between different bands and enable multi-band operations through on-channel tunneling. Transmission over multiple bands can happen simultaneously or not. Traffic can be distributed over single or multiple bands, and the MAC architecture for multi-band operations can be Independent, Distributed, or Unified according to the IEEE.~\cite{ieee4}. 
The MAC protocol usage in different bands varies based on the type of architecture employed. Independent architecture permits the use of different MAC addresses and protocols for each band, while information from upper layers is utilized to distribute traffic. Distributed architecture, however, assigns only one MAC address for each band and shares information at the local level, without the upper layer being aware of session transfer of same or different traffic between different bands. In unified MAC architecture, a new MAC layer is responsible for the dynamic transfer of traffic among different bands for concurrent or non-concurrent transmission. The current multi-band MAC protocol enables the transfer of traffic between different links, with this switching information being shared within the access point (AP). This process of transferring the same or different traffic to multiple bands is referred to as "multi-band switching". When multiple access points (APs) are used, switching traffic or transmitting multiple traffic concurrently or non-concurrently can cause issues for MAP-Co in the future. For example, if traffic is switched between multiple bands without using multi-band C-OFDMA, ICI can negatively impact the signal quality of the switched traffic. This occurs when signals from different communication channels interfere with the selected channel, resulting in signal degradation. To ensure optimal switch traffic, it is recommended to use a coordinated multi-band OFDM method to effectively select channels and avoid potential issues. Similarly, in the CSR, switching traffic to another band can create interference at the neighboring same band if the transmission power is not controlled in coordination with multi-band switching at another AP. In CBF, switching traffic to another band and using a beam for transmission can create interference if nulling is not performed to neighboring AP in the same band due to no coordination of multi-band switching. JTX by multi-band switching without MAP-Co can lead to interference and collision if the same channel/resource units are used, which can lower the throughput. When using single-band APs, the APs share channel information and determine parameters for using MAP-Co features. However, with multi-band APs, band selection occurs at the packet level, resulting in rapid switching of packets between multiple bands. However, sharing information among multi-band APs within the limited time frame imposed by fast switching and with the same packet arrival rate presents a significant challenge. One solution to this challenge is to use machine or deep learning to predict when multi-band switching will happen and notify other APs ahead of time. By doing this, APs can make better decisions in real-time and improve the efficiency of multi-band MAP-Co. Traffic prediction is a method for optimizing resource allocation across different bands, such as between ground stations, HAPS, and satellites, as proposed by~\cite{masaki}. Similarly, we predict traffic across multiple WLAN bands and schedule resources across these bands through MAP-Co accordingly.
\subsection{Future MAP-Co Architecture} 
This paper explores two types of MAP-Co architecture: MA-based and MC-based. The MC-based approach is centralized and uses a local server to connect all APs through a wired backhand network. In the future, we plan to introduce MAP-Co systems that use mmWave and THz WLAN, which will connect to the local server and all 2.4GHz APs, as shown in Fig.~\ref{future5}. However, the local server has limited computation power and communication capacity, so a semi-centralized architecture is necessary to balance the load at each MC and prevent congestion. To accomplish this, an intelligent policy for the semi-centralized MAP-Co architecture must be developed to determine the connection between multiple APs and MCs, as depicted in Fig.~\ref{future5}. The challenge for such a policy will be to consider the delay and dynamic nature of the WLAN environment, such as mobility and traffic variation. Therefore, we propose a deep-learning-based approach to determine MAP-Co control at each local server and how to distribute it among different local servers.

During our previous discussion, we covered MA-based architecture, which is a type of MAP-Co architecture that uses a semi-distributed approach. In this approach, one AP serves as the master, with the other APs acting as slaves. Although this approach holds promise, it poses several challenges and practical issues for future MAP-Co systems that include heterogeneous APs such as 2.4 GHz/60 GHz/THz. The current MAP-Co architecture only considers clusters of the same type of WLAN network, but the future MAP-based cluster will include other WLANs like mmWave and THz. This raises new questions, such as how to make a heterogeneous cluster that is dynamic and flexible based on the current network environment and connections, and what should be the optimal cluster size of the heterogeneous MAP-Co. Additionally, the existing MA-based architecture selects a fixed AP as the master, as shown in Fig.~\ref{future5}. This prompts the question of whether it is fair to always rely on one access point (AP) for all responsibilities, leading to potential overload. Alternatively, could a scalable MA-based architecture with higher distribution be a better solution? To address this issue, we can propose a dynamic cluster head approach that selects a master AP based on the overhead conditions and MAP-Co features of each AP. The master AP can be chosen from the same WLAN (intra-head) or a different WLAN (inter-head). However, checking each inter and intra cluster's AP as a master AP using a brute force algorithm is an NP-hard problem that cannot be done in real-time. Therefore, researchers need to propose an intelligent method to choose the intra or inter-cluster head or master AP based on traffic at each AP and WLAN conditions, maximizing network performance.

\section{Conclusion}
		\label{Conclusion}
The MAP-Co protocol has been considered for future WLANs, such as beyond-802.11be, beyond-802.11ay, and THz, due to its ability to solve issues related to ICI and resource utilization. However, most surveys and WLAN standards had focused on single AP communication features, such as MU-MIMO, OFDMA, and multi-link, without providing a comprehensive and organized state-of-the-art of the MAP-Co architecture and its features. 
In this article, we provided an organized state-of-the-art MAP-Co architecture and its features, which can be of two types: MC and MA-based coordinator. We discussed the advantages and drawbacks of each architecture. Additionally, we have reviewed MAP-Co features related to data transmissions, such as coordinated channel sounding, C-OFDMA, CSR, CBF, and JTX. 
We conducted a thorough analysis of several techniques for multi-AP channel sounding. Our review revealed that both explicit and implicit multi-AP channel sounding methods faced significant overhead challenges in future WLANs. To address this issue, we examined various solutions and found that certain methods, such as multiple component and codebook-based feedback or deep-learning-based channel sounding, effectively mitigated the overhead problems associated with multi-AP explicit and implicit channel sounding. 
Our research on C-OFDMA methods highlighted the necessity for a proactive and less time-consuming resource allocation approach. Additionally, we provided an overview of existing OBSS-PD, PCSR, and MAC-based CSR methods. We determined that many of these methods were reactive and caused significant delays before performing CSR, which may not meet the demands of upcoming applications. 
We conducted a detailed analysis of the latest developments in CBF and JTX and learned from it. Additionally, we shared potential advancements for MAP-Co in the realm of emerging WLAN, including blockage-aware MAP-Co, proactive multi-AP channel sounding, multi-band MAP-Co, and future MAP-Co architectures. For each of these areas, we identified open issues and provided suggestions on how to address them.

 \section*{Acknowledgment}
These research results were obtained from the commissioned research (No. 22403) by the National Institute of Information and Communications Technology (NICT), Japan, and through work supported by the National Science Foundation of the USA under Award No. 2210252. Additionally, this work is also carried out as part of “Research and Development of Ultra-Large Capacity Wireless LAN using Terahertz Waves”, supported by the Ministry of Internal Affairs and Communications (MIC), Japan.

\begin{IEEEbiography}
	[{\includegraphics[width=1in,height=1.25in,clip,keepaspectratio]{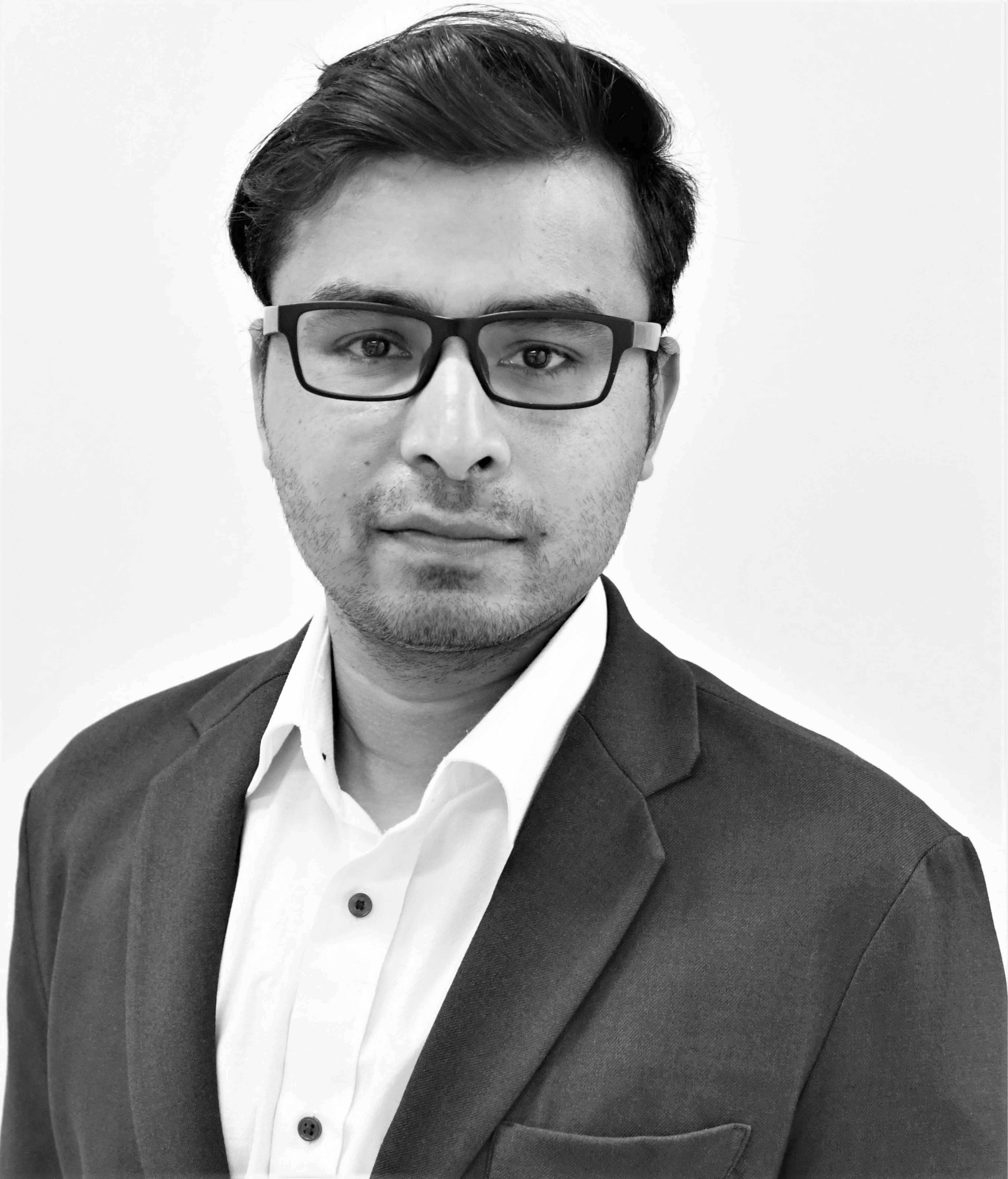}}]
	{Shikhar Verma} (Member, IEEE)
  received the M.Sc. and Ph.D. degrees from the Graduate School of Information Sciences (GSIS), Tohoku University, Sendai, Japan, in 2018 and 2021, respectively. Since 2021, he has been an Assistant Professor with GSIS, Tohoku University. Dr. Verma was also a recipient of the prestigious MEXT Scholarship and JSPS Fellowship. He also received the Dean’s Award from Tohoku University in 2021 and the Best Paper Award at IEEE ICC in 2018 and several other conferences. He serves as an Editor for the IEEE Internet of Things. 
\end{IEEEbiography}
\begin{IEEEbiography}
	[{\includegraphics[width=1in,height=1.5in,clip,keepaspectratio]{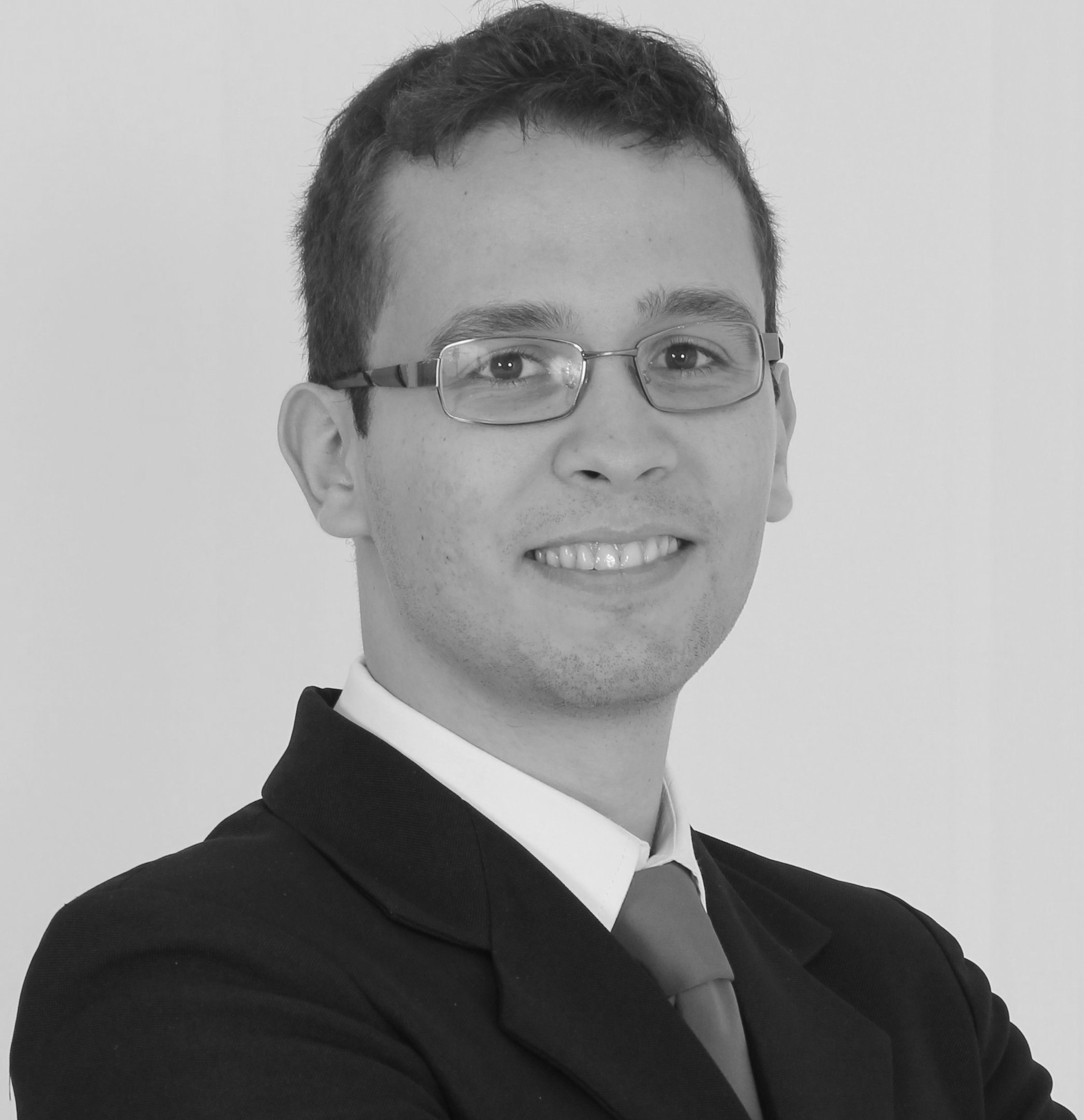}}]
	{Tiago Koketsu Rodrigues}
(Member, IEEE) previously Tiago Gama Rodrigues, is currently an Assistant Professor at Tohoku University. His research interests include artificial intelligence, machine learning, network modeling and simulation, and cloud systems. From 2017 to 2020, he was the Lead System Administrator of the IEEE Transactions on Vehicular Technology, overviewing the review process of all submissions and the submission system as a whole. He serves as an Editor for multiple IEEE journals. 

\end{IEEEbiography}
\begin{IEEEbiography}
	[{\includegraphics[width=1in,height=1.25in,clip,keepaspectratio]{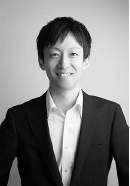}}]
	{Yuichi Kawamoto}  (Member, IEEE)
	is serving as an assistant professor at GSIS. He received his B.E. degree from the Information Engineering, and M.S. degree and Ph.D. degree from the Graduate School of Information Science (GSIS) at Tohoku University, Japan, in 2011 and 2013, 2016, respectively. He was a recipient of the prestigious Dean’s and President’s Awards from Tohoku University in March 2016. He has also received several best paper awards at conferences including IWCMC'13, GLOBECOM'13, and WCNC'2014. 
	
\end{IEEEbiography}
\begin{IEEEbiography}
	[{\includegraphics[width=1in,height=1.25in,clip,keepaspectratio]{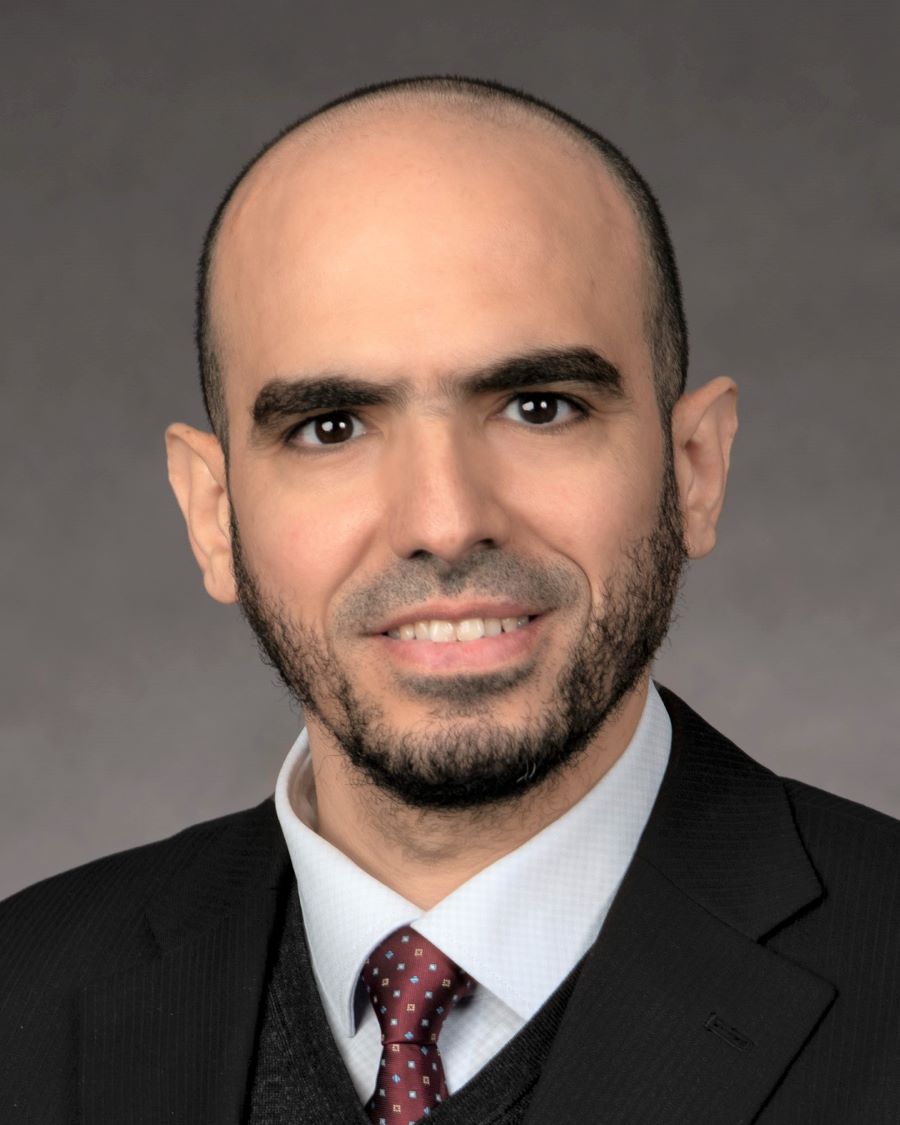}}]
	{Mostafa M. Fouda}  (Senior Member, IEEE) received the B.S. degree (as the valedictorian) and the M.S. degree B.S. (Hons.) and the M.S. degrees in Electrical Engineering from Benha University, Egypt, in 2002 and 2007, respectively, and the Ph.D. degree in Information Sciences from Tohoku University, Japan, in 2011. He is currently an Associate Professor with the Department of Electrical and Computer Engineering at Idaho State University, ID, USA. He also holds the position of a Full Professor at Benha University. He was an Assistant Professor at Tohoku University and a Postdoctoral Research Associate at Tennessee Technological University, TN, USA. He has (co)authored more than 210 technical publications. His current research focuses on cybersecurity, communication networks, signal processing, wireless mobile communications, smart healthcare, smart grids, AI, and IoT. He has guest-edited a number of special issues covering various emerging topics in communications, networking, and health analytics. He currently serves on the editorial board of the IEEE Internet of Things Journal (IoT-J), IEEE Transactions on Vehicular Technology (TVT), and IEEE Access. He has received several research grants, including NSF Japan-U.S. Network Opportunity 3 (JUNO3). 
\end{IEEEbiography}

\begin{IEEEbiography}
	[{\includegraphics[width=1in,height=1.25in,clip,keepaspectratio]{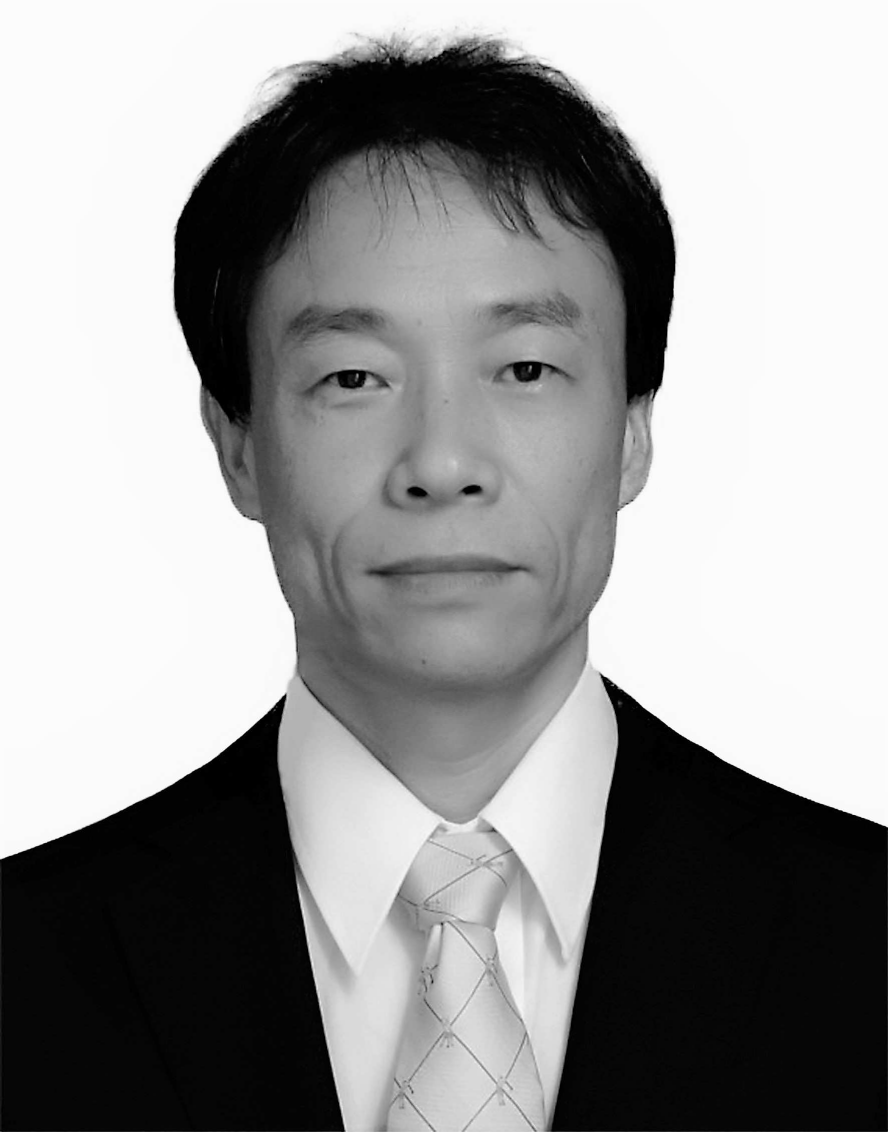}}]
	{Nei Kato}
 (Fellow, IEEE) is a full professor and the Dean with Graduate School of Information Sciences, Tohoku University, Japan. He has researched on computer networking, wireless mobile communications, satellite communications, ad hoc \& sensor \& mesh networks, UAV networks, AI, IoT, Big Data, and pattern recognition. He is the Editor-in-Chief of IEEE Internet of Things Journal. He served as the Vice-President (Member \& Global Activities) of IEEE Communications Society (2018-2021), and the Editor-in-Chief of IEEE Transactions on Vehicular Technology (2017-2021). He is a fellow of The Engineering Academy of Japan, a Fellow of IEEE, and a Fellow of IEICE.
\end{IEEEbiography}

\end{document}